\documentclass[10pt, pre,eps,twocolumn,showpacs, footnote]{revtex4-1}
\usepackage{url,ulem}
\usepackage{amsmath,amssymb}
\usepackage{fourier}
\newcommand{\sn}{\mathrm{sn}}

\usepackage{amsmath,amssymb}
\usepackage[dvipdfm]{graphicx}
\usepackage{color}
\usepackage{setspace}

\newcommand{\ave}[1]{\left\langle #1 \right\rangle}

\newcommand{\Mb}{\bar{M}}
\newcommand{\HA}{\mathcal{H}_{\rm A}}

\newtheorem{theorem}{Theorem}
\newtheorem{lemma}[theorem]{Lemma}

\newtheorem{remark}[theorem]{Remark}

\begin{document}
\title{ Nonlinear response for external field and perturbation
    in the Vlasov system }
\author{Shun Ogawa}
\email[E-mail: ]{sogawa@amp.i.kyoto-u.ac.jp}
\author{Yoshiyuki Y. Yamaguchi}
\affiliation{
	Department of Applied Mathematics and Physics, 
	Graduate School of Informatics, Kyoto University, 
	606-8501 Kyoto, Japan}
\begin{abstract}
    A nonlinear response theory is provided
    by use of the transient linearization method 
    in the spatially one-dimensional Vlasov systems.
    The theory inclusively gives responses
    to external fields and to perturbations
    for initial stationary states,
    and is applicable even to the critical point
    of a second order phase transition.
    We apply the theory to the Hamiltonian mean-field model,
    a toy model of a ferromagnetic body,
    and investigate the critical exponent associated with
    the response to the external field at the critical point
    in particular.
    The obtained critical exponent is nonclassical value $3/2$,
    while the classical value is $3$.
    However, interestingly, one scaling relation holds
    with another nonclassical critical exponent of susceptibility
    in the isolated Vlasov systems.
    Validity of the theory is numerically confirmed
    by directly simulating temporal evolutions of the Vlasov equation.
\end{abstract}
\maketitle 

\section{Introduction}
The Vlasov equation describes dynamics of Hamiltonian systems
with long-range interactions,
including self-gravitating systems and plasma systems,
in the limit of large number of particles
\cite{BH77,Dobrushin79,Spohn}.
The Vlasov system has continuously infinite number of stationary states,
and stable ones are called quasistationary states (QSS) \cite{YYY04},
whose life times diverge with increasing the number of particles.
The long-lasting QSS brings a problem
of understanding response to external field or perturbation.

In the Vlasov system,
the linear response theory for the external field 
has been recently developed \cite{AP12, SO12}. 
This theory gives the asymptotic value of
observable associated with the external field
if the field is small enough.
One remarkable product of the theory is
that the critical exponent of zero-field susceptibility
in a ferromagnetic model takes
the nonclassical value $1/4$
in an isolated system \cite{SO13},
while the classical mean-filed theory gives the exponent $1$
in an isothermal system.
However, the naive perturbation theory cannot work at the critical point
of a second order phase transition due to divergence of the linear response,
and hence the theory is not available to obtain
the critical exponent for response to the external field
at the critical point.
Another disadvantage of the linear response theory is,
obviously, that amplitude of the external field must be small enough.

Perturbation to a stable stationary state exponentially damps,
as shown by Landau \cite{LDL45}.
The exponential damping stands for the linear analysis of the Vlasov equation,
and nonlinear effects tend to stop the damping \cite{TMO65}.
Thus, there is a competition between the Landau damping
and nonlinear trapping effects. Based on this competition,
creations of small traveling clusters are discussed phenomenologically
\cite{JB09}.
The nonlinear effects make the response to perturbation non-trivial
even for stable stationary states.

Our goal is to construct a nonlinear response theory,
which inclusively describes responses
to the external field and to perturbation,
and which works well even just on the critical point.
The theory is constructed by using 
the transient linearization, which is called T-linearization,
method based on the asymptotic-transient decomposition
of the distribution function \cite{CL98, CL03, CL09}.
This method is originally proposed to investigate
dynamical asymptotic states given by nonlinear superposition
of Bernstein-Greene-Kruskal solutions \cite{MB94}. 
We apply the method to a simple toy model,
the Hamiltonian mean-field (HMF) model \cite{InagakiKonishi93,MA95},
in order to capture asymptotically stationary states
starting from stationary states with the external field 
and perturbations.
The main consequence of this method is that
the asymptotic states are equivalent to rearrangement of
initial distribution functions
along the energy contours associated with the asymptotic
effective Hamiltonian. 
Accordingly, this procedure induces a self-consistent equation
for the asymptotic state.

We emphasize progresses by the present article:
The self-consistent equation is expanded with respect to a small
order parameter in order to analyze the scaling relations.
The expansion method is different with and might be simpler than
one of the previous works \cite{CL03, CL09}.
Thanks to the expansion, we obtain one nonclassical critical exponent
for response to the external field at the critical point.
Interestingly, the nonclassical critical exponent
satisfies a classical scaling relation with the previously mentioned
nonclassical critical exponent for the zero-field susceptibility.
The theoretical predictions are quantitatively examined
by direct numerical simulations of the Vlasov equation.
A similar idea of the rearrangement is presented in Ref. \cite{PdB11},
but there is no theoretical justification
as the nonlinear trapping and the T-linearization,
and hence limitation of the theory was not clear.
We clarify hypotheses to ensure validity of the theory,
and, as a result, we can discuss the origins of discrepancies
which will be observed between the theory and the numerical
tests for response to perturbation.

We mention other previous studies on nonlinear dynamics 
of the Vlasov equation.
A bifurcation from spatially homogeneous to inhomogeneous states
is investigated by constructing unstable manifolds
of the unstable homogeneous stationary states 
in the weak instability limit 
for the Vlasov-Poisson equation \cite{JDC95}.
The theory predicts that inhomogeneity in asymptotic states
increases as a quadratic function of the distance
from the critical point in a parameter space, like temperature.
This prediction is numerically confirmed in a one-dimensional (1D) 
self-gravitating system \cite{AVI05}.
The theory provided in the present article also reproduces
the quadratic scaling successfully.

The organization of present article is as follows. 
We introduce the spatially 1D and periodic Vlasov equation
and the T-linearization method \cite{CL03, CL09} in Sec.~\ref{sec:TL}.
The method is exhibited in Ref.~\cite{CL09} in details, 
but it might be worthwhile to rephrase the derivation of T-linearization 
in a simpler form with confirming the necessary hypotheses.
In Sec.~\ref{sec:HMF},
the general theory is applied to the HMF model,
and we expand the asymptotically self-consistent equation
with respect to the small order parameter. 
Theoretical consequences are arranged in Sec.~\ref{sec:theorical-consequences}
with the aid of the expanded self-consistent equation,
and these predictions are numerically examined in Sec.~\ref{sec:numerics}.
Conclusion and remarks are in Sec.~\ref{sec:conclusion}.

\section{Transient Linearization in Vlasov system}
\label{sec:TL}
We consider spatially 1D and 2$\pi$-periodic Hamiltonian systems
described by the Hamiltonian
\begin{equation}
	\label{eq:Nbody-Hamiltonian}
	H_{N} = 
	\sum_{i = 1}^N\frac{p_i^2}{2} + \frac{1}{2N}\sum_{i,j=1}^N V(q_i -q_j)
        + \sum_{i=1}^N U (q_i),
\end{equation}
where $q_{i}\in (-\pi,\pi]$ is the position of $i$-th particle
for $i = 1,2,\cdots,N$, 
$p_{i}\in\mathbb{R}$ the conjugate momentum,
and the interaction $V(q)$ is even.
The external field is represented
by the on-site potential $U$.
In the large limit of $N$, the evolution of this system  
is well described 
in terms of the single body distribution $f(q,p,t)$
governed by the Vlasov equation \cite{BH77,Dobrushin79,Spohn}: 
\begin{equation}
	\label{eq:vla}
	\frac{\partial f}{\partial t} + \{\mathcal{H}[f], f \} = 0.
\end{equation}
The effective Hamiltonian $\mathcal{H}[f]$ is given by 
\begin{equation}
	\label{eq:vla2}
	\begin{split}
		\mathcal{H}[f] &= \frac{p^2}{2} + \mathcal{V}[f](q,t) + U(q), \\
		\mathcal{V}[f] &=
		\int_{-\pi}^{\pi} dq' \int_{-\infty}^{\infty} V(q-q') f(q',p',t)dp'
	\end{split}
\end{equation}
and the Poisson bracket $\{a,b\}$ is defined by
\begin{equation}
	\{a, b\} \equiv \frac{\partial a}{\partial p}\frac{\partial b}{\partial q}
	-\frac{\partial a}{\partial q}\frac{\partial b}{\partial p}.
\end{equation}

\subsection{Asymptotic-transient  decomposition}
We take the initial condition of a stationary state
with a perturbation as
\begin{equation}
    f_{0}(q,p) = f_{\rm ini}(q,p) + \epsilon g(q,p).
\end{equation}
We refer to $f_{0}$ and $f_{\rm ini}$ as
the initial state and the initial stationary state respectively.
The perturbation must satisfy
\begin{equation}
    \label{eq:intg0}
    \iint_{\mu} g(q,p) dqdp = 0
\end{equation}
to keep the normalization condition,
where $\mu$ represents the whole $(q,p)$ plane.
The Vlasov equation \eqref{eq:vla} evolves the initial state $f_{0}(q,p)$
to $f(q,p,t)$ at time $t$.
We introduce a hypothesis on the asymptotic state:
 \begin{enumerate}
       \item[{\bf H0}]
     The state $f$ goes to a stationary state $f_{\rm A}$
     asymptotically. 
 \end{enumerate}
The hypothesis {\bf H0} is the basic hypothesis of the present theory.

The asymptotic state $f_{\rm A}$ is rigorously defined
by the Bohr transform defined as
\begin{equation}
    \mathcal{B}_{\omega} [a] \equiv 
    \lim_{\sigma \to \infty} \frac{1}{\sigma} \int_0^\sigma a(t) e^{-i\omega t}dt. 
\end{equation}
This transform picks up the asymptotic oscillating mode of $a(t)$
with frequency $\omega$.
The asymptotic stationary state $f_{\rm A}$ has no oscillating modes,
thus we define it as
\begin{equation}
    \label{eq:fA}
    f_{\rm A}(q,p) = \mathcal{B}_{0} [f]. 
\end{equation}
Validity of this definition is guaranteed by the equality
\begin{equation}
    \mathcal{B}_{0}[f] = \lim_{t\to\infty} f.
\end{equation}
See Appendix \ref{app:B0} to derive it.

For later convenience, 
we decompose $f(q,p,t)$ into the initial state $f_{0}(q,p)$ 
and evolving perturbation as 
\begin{equation}
 \label{eq:fge}
	f(q,p,t) = f_0(q,p) + \epsilon g_{\rm E}(q,p,t).
\end{equation}
The second hypothesis is:
\begin{enumerate}
      \item[{\bf H1}]
    The evolving part $\epsilon g_{\rm E}$ is of $O(\epsilon)$.
\end{enumerate}
The hypothesis {\bf H1} implies that the state $f$ is
in a $O(\epsilon)$ neighborhood of the initial state $f_{0}$.
Thus, we decompose $f(q,p,t)$ into the asymptotically
stationary surviving part $f_{\rm A}$
and the transient part vanishing at $t \to \infty$ as
\begin{equation}
    \label{eq:fgt}
    f(q,p,t) = f_{\rm A}(q,p) + \epsilon g_{\rm T}(q,p,t),
\end{equation}
where the transient part $\epsilon g_{\rm T}$ is of $O(\epsilon)$.
The hypothesis {\bf H1} and one of its consequences \eqref{eq:fgt}
are used to perform the T-linearization.

The decomposition of $f$, Eq.~\eqref{eq:fgt}, induces the decomposition
of the effective Hamiltonian as
\begin{equation}
    \label{eq:H-decompose}
    \mathcal{H}[f](q,p,t)
    = \mathcal{H}_{\rm A}(q,p) + \epsilon \mathcal{V}_{\rm T}(q,p,t),
\end{equation}
where
\begin{equation}
    \mathcal{H}_{\rm A}(q,p) = \dfrac{p^{2}}{2} + \mathcal{V}_{\rm A}(q)
    + U(q),
\end{equation}
and 
\begin{equation}
    \mathcal{V}_{\rm A}(q) = \mathcal{V}[f_{\rm A}],
    \qquad
    \mathcal{V}_{\rm T}(q,t) = \mathcal{V}[g_{\rm T}].
\end{equation}
Thanks to the Jeans theorem \cite{Jeans15},
we may assume that the asymptotic distribution $f_{\rm A}$,
which is stationary, is a function of the asymptotic Hamiltonian
$\mathcal{H}_{\rm A}$, and the Hamiltonian $\mathcal{H}_{\rm A}$
is determined by $f_{\rm A}$. The asymptotic distribution
must be therefore determined self-consistently.

\subsection{T-linearized Vlasov equation}
The T-linearization is performed by omitting $O(\epsilon^{2})$ terms:
\begin{enumerate}
      \item[{\bf H2}] We omit $O(\epsilon^{2})$ terms.
\end{enumerate}
Substituting Eqs.~\eqref{eq:fge} and \eqref{eq:H-decompose} into 
the Vlasov equation \eqref{eq:vla} and
using the hypothesis {\bf H2},
we have the T-linearized Vlasov equation,
\begin{equation}
    \label{eq:TLVlasov} 
    \frac{\partial f}{\partial t} + 
    \left\{\mathcal{H}_{\rm A}, f\right\}
    + \epsilon \left\{\mathcal{V}_{\rm T},  f_0\right\}= 0. 
\end{equation}
We stress that the term including the asymptotic Hamiltonian
$\mathcal{H}_{\rm A}$ still remains nonlinear 
although this equation is called T-``linearized'' equation.
\begin{remark} 
Following the previous works \cite{CL98, CL03, CL09},
we substituted the decomposition \eqref{eq:fge}
instead of Eq.~\eqref{eq:fgt}.
For the setting discussed in the present article,
we may replace $f_{0}$ with $f_{\rm A}$
if they are even with respect to $p$,
since contribution from this term vanishes as shown later.
\end{remark}
By introducing the operator $\mathcal{L}_{\rm A}$ as 
\begin{equation}
    \label{eq:L-A}
    \mathcal{L}_{\rm A} [\bullet] =
    -
    \left\{\mathcal{H}_{\rm A}, \bullet \right\},
\end{equation}
the T-linearized Vlasov equation \eqref{eq:TLVlasov} is written in the form 
\begin{equation}
	\begin{split}
	\frac{\partial f}{\partial t} &= 
	\mathcal{L}_{\rm A}[f] - \epsilon \left\{\mathcal{V}_{\rm T},
          f_{0}\right\}.  
          \end{split}
\end{equation}
The formal solution to T-linearized Vlasov equation with
the initial condition $f(q,p,0) = f_0(q,p)$ is \cite{SO12}
\begin{equation}
	\label{eq:solA0}
	f(q,p,t) = e^{t \mathcal{L}_{\rm A}} f_{0}(q,p) 
	- \epsilon \int_{0}^{t} e^{(t-s)\mathcal{L}_{\rm A}}
        \left\{\mathcal{V}_{\rm T}(q,s),  f_{0}\right\}ds.
\end{equation}
The first and second terms of the right-hand-side
are called the O' Neil term and the Landau term
and are denoted by $f_{\rm O}$ and $\epsilon f_{\rm L}$
respectively.
In the next subsection,
we will give a simple expression of the self-consistent equation
to determine the asymptotic distribution $f_{\rm A}$
with the aid of the ergodic like formula.

\subsection{Ergodic like formula}
\label{sec:ergod}
The operator $\exp(t\mathcal{L}_{\rm A})$
in Eq. \eqref{eq:solA0} drives a function $\psi(q,p,0)$ to $\psi(q,p,t)
=\psi(q(-t),p(-t),0)$,
where $(q(t),p(t))$ is the solution
to the Hamiltonian equation of motion associated with
the asymptotic Hamiltonian $\HA$ with the initial point $(q,p)$.
Thus, the Bohr transform
\begin{equation}
    \mathcal{B}_{0}\left[ \exp(t\mathcal{L}_{\rm A})\psi(q,p) \right]
    = \lim_{\sigma \to \infty} \frac{1}{\sigma} \int_{0}^{\sigma}
    e^{t\mathcal{L}_{\rm A}}\psi(q,p)dt
\end{equation}
is read as the time average
of $\psi(q,p)$ along the orbit $(q(-t),p(-t))$.
The ergodic like formula replaces the time average
with the partial phase space average under constraints of the integral \cite{CL03, CL09}.

The asymptotic Hamiltonian $\mathcal{H}_{\rm A}$
is integrable,
since it is stationary and is a spatially 1D system.
We can hence introduce the angle-action variables $(\theta,J)$,
whose temporal evolutions for the initial point $(\theta,J)$ are
\begin{equation}
    \theta(t) = \theta + \Omega(J)t, \quad
    J(t) = J,
\end{equation}
where $\Omega(J)={\rm d}\mathcal{H}_{\rm A}/{\rm d}J$.
The time average on the energy contour,
which is equivalent with the iso-$J$ line,
is thus obtained by the average over $\theta$.
The ergodic like formula is expressed by
\begin{equation}
    \label{eq:ergodic}
    \mathcal{B}_{0}\left[ \exp(t\mathcal{L}_{\rm A})\psi(q,p) \right]
    = \ave{\psi}_{J},
\end{equation}
where the partial phase space average is defined by
\begin{equation}
    \ave{\psi}_{J} = \dfrac{1}{2\pi} \int_{-\pi}^{\pi}
    \psi (q(\theta,J),p(\theta,J)) d\theta.
\end{equation}
The subscript $J$ in the left-hand-side represents
that the partial phase space average is taken on the iso-$J$ line.

We will derive a simple expression of the asymptotic distribution $f_{\rm A}$
by using the ergodic like formula \eqref{eq:ergodic}.
From the definition \eqref{eq:fA},
the asymptotic distribution $f_{\rm A}$ is decomposed as
\begin{equation}
    f_{\rm A} = \mathcal{B}_{0}[f_{\rm O}] + \epsilon \mathcal{B}_{0}[f_{\rm L}].
\end{equation}
Contribution from the O'Neil term is written in the form
\begin{equation}
    \label{eq:fA-ONeil}
    \mathcal{B}_{0}[f_{\rm O}]
    = \mathcal{B}_{0}\left[
      e^{t \mathcal{L}_{\rm A}} f_0(q,p) \right]
    = \ave{f_{0}}_{J}.
\end{equation}
Here after, we put an additional hypothesis: 
\begin{enumerate}
	\item[{\bf H3}] 
      The initial state $f_0(q,p)$ is even with respect to $p$. 
\end{enumerate}
\begin{lemma}
	\label{lem:L-vanish}
  	Contribution to $\mathcal{H}_{\rm A}$ from
      the Landau term vanishes,
      that is $\mathcal{V}\left[\mathcal{B}_{0}[f_{\rm L}]\right]=0$,
      under the hypotheses from {\bf H0} to {\bf H3}.
\end{lemma}
{\bf Proof:} 
Thanks to the periodicity for $q$, 
we perform Fourier transform of the force field 
$-\partial \mathcal{V}_{\rm T}/\partial q$ with respect to $q$ as 
\begin{equation}
	-\dfrac{\partial \mathcal{V}_{\rm T}}{\partial q}(q,s) 
	= \sum_{k \in \mathbb{Z}}\mathcal{T}_{k}(s) e^{ikq}.	
\end{equation}
We note that 
$\mathcal{T}_{k}(t)$ converges to $0$ as $t \to \infty$ rapidly 
so that $\int_0^\infty|\mathcal{T}_{k}(t)| dt < \infty$. 
This is because the transient force field $\mathcal{T}(s)$ represents 
the term damping exponentially \cite{LDL45} or algebraically \cite{BOY11}, 
thanks to the hypothesis {\bf H0}. 
It has been reported that the algebraic damping is equal to or faster than the inverse square of time, $t^{-2}$ 
for spatially 1D systems, 
so that the $\mathcal{T}_k(s)$ is to be an $L^1$ function. 
Substituting it into the Landau part, we have
\begin{equation}
	\begin{split}
		&\mathcal{B}_{0}[ f_{\rm L}]
		=  \sum_{k \in \mathbb{Z}} B_{k}, \\
		&B_k 
		\equiv  -\lim_{\sigma\to\infty} \dfrac{1}{\sigma} \int_{0}^{\sigma} dt
		 \int_{0}^{t} e^{(t-s)\mathcal{L}_{\rm A}} 
		  \mathcal{T}_{k}(s) e^{ikq} \frac{\partial f_0}{\partial p}(q,p) ds. 
	\end{split}
\end{equation}
It is possible to replace the upper bound of $s$ with $\infty$ by adding a vanishing part, 
since the integrand except for $\mathcal{T}_k(s)$ is bounded, and 
$\mathcal{B}_0\left[\int_t^\infty \mathcal{T}_k(s) ds \right] = 0$
with the aid of Eq. \eqref{eq:Frechet} \cite{CL03, CL09}. 
The $B_{k}$ is modified as 
\begin{equation}
	\begin{split}
		B_k &= 
		 - \int_{0}^{\infty}ds \mathcal{T}_{k}(s) 
		  \lim_{\sigma\to\infty} \dfrac{1}{\sigma} \int_{-s}^{\sigma-s}
		   e^{u\mathcal{L}_{\rm A}} 
		  e^{ikq} \frac{\partial f_0}{\partial p}(q,p) du  , 
	\end{split}
\end{equation}
where we have changed the variable from $t$ to $u=t-s$.
Further, 
by use of the ergodic like formula \eqref{eq:ergodic} for 
integration with respect to $u$, 
we obtain 
\begin{equation}
	\lim_{\sigma\to\infty} \dfrac{1}{\sigma} \int_{-s}^{\sigma-s}
		   e^{u\mathcal{L}_{\rm A}} 
		  e^{ikq} \frac{\partial f_0}{\partial p}(q,p) du~
		  = \left\langle e^{ikq} \frac{\partial f_0}{\partial p}(q,p) \right \rangle_J,
\end{equation}
for each $s$.
Then, we have 
\begin{equation}
	\label{eq:BfL}
	B_k = -\left \langle e^{ikq} \frac{\partial f_0}{\partial p}(q,p) \right \rangle_J 
	 \int_{0}^{\infty} \mathcal{T}_{k}(s) ds.
\end{equation}
The action variable is determined by the asymptotic Hamiltonian
$\HA$ which is even with respect to $p$,
and hence the iso-$J$ line is symmetric under the transform $p\mapsto -p$.
Remembering that $f_{0}$ is even with respect to $p$,
we conclude
 $\mathcal{V}\left[\mathcal{B}_{0}[f_{\rm L}]\right]=0$.~$\blacksquare$\\

We put a physical interpretation of 
{\bf Lemma \ref{lem:L-vanish}}
as the following {\bf Remark \ref{remark}}: 
\begin{remark}
    \label{remark}
    Under the introduced hypotheses,
    we may omit the transient potential $\mathcal{V}_{\rm T}$
    in the T-linearized Vlasov equation \eqref{eq:TLVlasov},
    and may neglect the transient temporal evolution
    including the Landau damping \cite{CL98}.
    In other words, the nonlinear trapping effect
    dominates the Landau damping
    and the state is rapidly trapped at the asymptotic state.
    Such a situation is suitable around the stable side
    of a stability threshold,
    where the Landau damping rate is almost zero.
\end{remark}

Contribution from the O'Neil term \eqref{eq:fA-ONeil}
and {\bf Lemma \ref{lem:L-vanish}} give 
a simple expression of the asymptotic effective Hamiltonian $\mathcal{H}_{\rm A}$ as 
\begin{equation}
    \label{eq:fA-avef0}
   \mathcal{H}_{\rm A}(q,p)= \mathcal{H}\left[f_{\rm A}(q,p)\right]= \mathcal{H}\left[\ave{f_{0}}_{J}\right].
\end{equation}
We note that the above equation is a self-consistent equation,
since the right-hand-side depends on the asymptotic effective Hamiltonian $\mathcal{H}_{\rm A}$ 
through the definition of the angle-action variables $(\theta,J)$.
The equation \eqref{eq:fA-avef0} is physically interpreted
as the rearrangement of $f_{0}$ to make $f_{\rm A}$ constant
along the iso-$J$ line \cite{PdB11}.
In the next section, we apply the general theory described
in this section to the HMF model.

\section{Application to the Hamiltonian mean-field model}
\label{sec:HMF}
Let us consider the HMF model,
which is a simple toy model of a ferromagnetic body.
The interaction is written by $V(q) = -\cos q$,
and the on-site potential is $U(q) = -h_{x} \cos q-h_{y}\sin q$
in Eq. \eqref{eq:Nbody-Hamiltonian},
where $U(q)$ represents potential by the external magnetic field
$(h_{x},h_{y})$.
The effective Hamiltonian is written as
\begin{equation}
    \mathcal{H}[f](q,p,t)
    = \dfrac{p^{2}}{2} - (\mathcal{M}_{x}[f] +h_{x})\cos q
    - (\mathcal{M}_{y}[f] +h_{y})\sin q,
\end{equation}
where $(\mathcal{M}_{x}[f],\mathcal{M}_{y}[f])$ is the magnetization vector
defined by
\begin{equation}
    \label{eq:Mx}
    \mathcal{M}_{x}[f] = \iint_{\mu} \cos q~ f(q,p,t) dqdp, 
\end{equation}
and
\begin{equation}
    \mathcal{M}_{y}[f] = \iint_{\mu} \sin q~ f(q,p,t) dqdp. 
\end{equation}

For simplicity, we consider the following situation,
mentioned as another hypothesis:
\begin{enumerate}
      \item[{\bf H4}]
    The magnetization vector and the external magnetic field
    asymptotically point to the positive $x$-direction.
\end{enumerate}
In other words, the asymptotic Hamiltonian $\HA$ is written as
\begin{equation}
    \label{eq:HA-HMF}
    \HA(q,p) = \dfrac{p^{2}}{2} - (M+h)\cos q,
\end{equation}
where $M$ and $h$ are the values of the asymptotic magnetization
and the external magnetic field respectively and they are positive constants.
We note that the initial distribution $f_{\rm ini}$ and perturbation
$g$ must be suitably chosen to fit this situation.

\subsection{The asymptotically self-consistent equation}
\label{sec:self-consistent-HMF}

The asymptotically self-consistent equation \eqref{eq:fA-avef0}
implies the equation for $M$ as
\begin{equation}
    \label{eq:A-SC}
    M = \iint_{\mu} \cos q \ave{f_{0}}_{J} dqdp.
\end{equation}
It is worth noting that the above self-consistent equation
has another expression of
\begin{equation}
    \label{eq:A-SC-2}
    M = \iint_{\mu} \ave{\cos q}_{J} f_{0}(q,p) dqdp.
\end{equation}
Indeed, we can show the following Lemma \ref{lem:j-move}:

\begin{lemma}
\label{lem:j-move}
Let $A(q,p)$ and $B(q,p)$ be functions on $(q,p)$ plane,
and both $A\ave{B}_{J}$ and $\ave{A}_{J}B$ are integrable
in a rectangle region $(-\pi,\pi] \times I_{J}\ni (\theta,J)$.
Let $R_{J}$ be the corresponding region on the $(q,p)$ plane
with the rectangle region. 
Then, the following equality holds:
\begin{equation}
    \iint_{R_{J}} A\ave{B}_{J} dqdp = \iint_{R_{J}} \ave{A}_{J}B ~dqdp.
\end{equation}
\end{lemma}
{\bf Proof:} The transform $(q,p)\mapsto (\theta,J)$ is canonical,
and hence $dqdp=d\theta dJ$. Using the integrability of
$A\ave{B}_{J}$ and $\ave{A}_{J}B$, we have
\begin{equation}
    \begin{split}
         &\iint_{R_{J}} A\ave{B}_{J} dqdp
        = \int_{I_{J}} \left( \int_{-\pi}^{\pi} A\ave{B}_{J} d\theta \right) dJ\\
        &= 2\pi \int_{I_{J}} \ave{A}_{J} \ave{B}_{J} dJ 
        = \int_{I_{J}} \left( \int_{-\pi}^{\pi} \ave{A}_{J} B d\theta \right) dJ\\
        &= \iint_{R_{J}} \ave{A}_{J} B ~dqdp.
        \quad \blacksquare
    \end{split}
\end{equation}

\begin{figure}
    \centering
    \includegraphics[width=8cm]{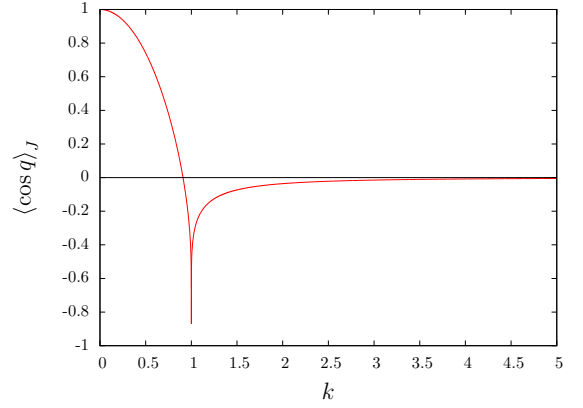}
    \caption{$\ave{\cos q}_{J}$ as the function of $k$.
      The value of $\ave{\cos q}_{J}$ is $-1$ in the limit $k\to 1$,
      since $K(k)$ diverges while $E(k)\to 1$.}
    \label{fig:cosq_J}
\end{figure}

With the aid of the concrete forms of angle-action variables
$(\theta,J)$ in the HMF model \cite{BOY10},
we have $\ave{\cos q}_{J}$ as
\begin{equation}
    \label{eq:cosqJ-of-k}
    \ave{\cos q}_{J}
    = \left\{
      \begin{array}{ll}
          \dfrac{2E(k)}{K(k)} - 1, & k<1 \\ 
          & \\
          \dfrac{2k^{2}E(1/k)}{K(1/k)} - 2k^{2}+1, & k>1 \\ 
      \end{array}
    \right.
\end{equation}
where the functions $K(k)$ and $E(k)$ are the complete elliptic integrals of
the 1st and the 2nd kinds respectively.
The modulus $k$ is defined by
\begin{equation}
    \label{eq:k-p}
    k = \sqrt{\dfrac{p^{2}+2\Mb(1-\cos q)}{4\Mb}},
\end{equation}
where $\Mb = M + h$.
The intervals $k<1$ and $k>1$ imply inside and outside separatrix respectively.
See Appendix \ref{app:elliptic} for derivations of $\ave{\cos q}_{J}$,
and Fig.~\ref{fig:cosq_J} for its graphical presentation.
We know $k$ for a given point $(q,p)$, thus the asymptotically self-consistent
equation \eqref{eq:A-SC-2} can be solved numerically at least.

In the followings, we theoretically analyze the asymptotically self-consistent
equation \eqref{eq:A-SC-2} for small $M$ case.
One advantage of this theoretical treatment is 
that we can obtain the critical exponent $\delta$,
which is, at the critical point of the second order phase transition,
defined as
\begin{equation}
    \label{eq:critical-exponent-delta}
    M \propto h^{1/\delta}.
\end{equation}
The exponent $\delta$ is $3$ in the classical mean-field theory \cite{HN11},
but we will show that the exponent $\delta$ is $3/2$
in the present isolated Vlasov system.

\subsection{Assumptions and expanded self-consistent equation}
We introduce some assumptions to derive the theoretical approximation
of the asymptotically self-consistent equation \eqref{eq:A-SC-2}.
\begin{enumerate}
      \item[{\bf A0}] The asymptotic magnetization $M$ is small enough.
\end{enumerate}
This assumption suggests that $h$ is also small enough,
and permits to expand the right-hand-side of \eqref{eq:A-SC-2}
in a power series of $\Mb$.
Another assumption is for the initial distribution $f_{0}$:

\begin{enumerate}
      \item[{\bf A1}] $f_{0}(q,p)$ satisfies the hypothesis {\bf H3},
      is smooth with respect to $p$, 
      and is bounded on $(-\pi, \pi] \times \mathbb{R}$.
\end{enumerate}
The assumption ${\bf A1}$ induces the following {\bf Lemma \ref{lem:f-bd}}.

\begin{lemma}
\label{lem:f-bd} 
Let $f$ satisfy the assumption {\bf A1}.
Then, there exists a positive constant $c$ such that
\begin{equation}
    \label{eq:assumption-2}
    \left| f(q,p) - f(q,0) \right| < cp^{2}
\end{equation}
holds for any $(q,p)\in (-\pi,\pi]\times\mathbb{R}$.
\end{lemma}
See Appendix \ref{app:lemma2} for the proof.

Under the above assumptions,
we will show that the asymptotically self-consistent equation \eqref{eq:A-SC-2}
is expanded as:
\begin{equation}
    \label{eq:A-SC-expansion}
    M = L_{1/2}[f_{0}] \Mb^{1/2} + L_{1}[f_{0}] \Mb + L_{3/2}[f_{0}] \Mb^{3/2} + O(\Mb^{7/4}),
\end{equation}
where
\begin{equation}
    \begin{split}
        & L_{1/2}[f_{0}] = \Mb^{-1/2} \iint_{\mu} f_{0}(q,0) \ave{\cos q}_{J} dqdp \\
        & L_{1}[f_{0}] = - \dfrac{1}{2} \iint_{\mu} \dfrac{f_{0}^{(1)}(q,p)}{p}dqdp \\
        & L_{3/2}[f_{0}] = \Mb^{-3/2} \iint_{\mu} f_{0}^{(2)}(q,0)
        \left( \dfrac{p^{2}}{2}\ave{\cos q}_{J} + \dfrac{\Mb}{4} \right) dqdp.
    \end{split}
\end{equation}
We introduced the symbol $f_{0}^{(n)}$ for the $n$-th partial derivative
of $f_{0}$ with respect to $p$.
The coefficients $L_{1/2}$ and $L_{3/2}$ have the prefactor
depending of $\Mb$, but are of $O(1)$.
Indeed, changing variables from $(q,p)$ to $(q,k)$, we have
\begin{equation}
    \begin{split}
        L_{1/2}
        & = 8 \left( \int_{0}^{1} dk \int_{0}^{q_{\max}} dq~
          + \int_{1}^{\infty} dk \int_{0}^{\pi} dq~ \right)
        \psi_{1/2}(q,k)
    \end{split}
\end{equation}
and
\begin{equation}
    \begin{split}
        L_{3/2}
        & = 8 \left( \int_{0}^{1} dk \int_{0}^{q_{\max}} dq~
          + \int_{1}^{\infty} dk \int_{0}^{\pi} dq~ \right)
        \psi_{3/2}(q,k)
           \end{split}
\end{equation}
with
\begin{equation}
    q_{\max} = \cos^{-1}(1-2k^{2}) = 2\sin^{-1}k.
\end{equation}
The integrands are
\begin{equation}
    \psi_{1/2}(q,k)
    = \dfrac{f_{0}(q,0)\ave{\cos q}_{J}}{\sqrt{1-k^{-2}\sin^{2}(q/2)}}
\end{equation}
and
\begin{equation}
    \psi_{3/2}(q,k) =
    \dfrac{f_{0}^{(2)}(q,0) \left[ \left(
          2k^{2}-1+\cos q \right) \ave{\cos q}_{J}+1/4\right]}
    {\sqrt{1-k^{-2}\sin^{2}(q/2)}}.
\end{equation}
We used the relation \eqref{eq:k-p}
for getting the function $\psi_{3/2}$.

\subsection{Expansion of the self-consistent equation}
\begin{figure}[tb]
	\centering
	\includegraphics[width=8cm]{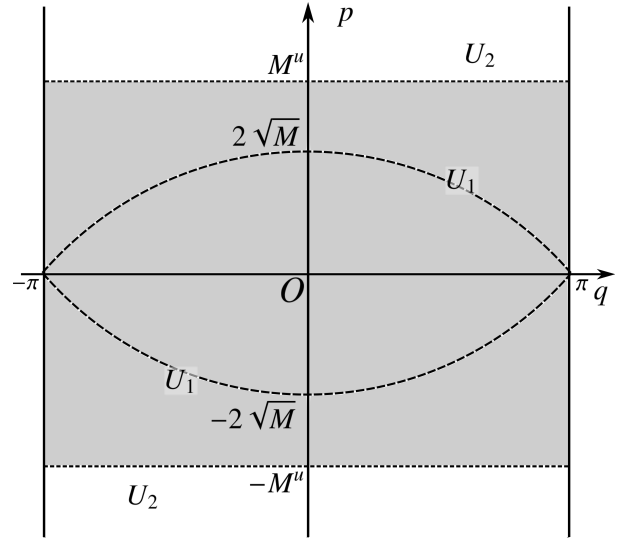}\\
	\caption{
          Schematic picture of division of the $\mu$ space.
          The region $U_{1}$ consists of
 	gray region,
          and $U_{2}$ of white region. 
          The broken curve in the region $U_{1}$ is a separatrix.}
 	\label{fig:muspace}
\end{figure}

Let us expand the right-hand-side of
the asymptotically self-consistent equation \eqref{eq:A-SC-2},
which is denoted by
\begin{equation}
    I = \iint_{\mu} \ave{\cos q}_{J} f_{0}(q,p) dqdp.
\end{equation}
The basic strategy is to divide the whole $\mu$-space
into the two parts $U_{1}$ and $U_{2}$, where
\begin{equation}
    U_{1} = \{ (q,p) ~|~ |p|<p_{\ast} \},
    \quad
    U_{2} = \{ (q,p) ~|~ |p|>p_{\ast} \}.
\end{equation}
The boundary $p_{\ast}=M^{u}$ between $U_{1}$ and $U_{2}$
is determined to satisfy the following requirements:
\begin{itemize}
      \item In $U_{1}$, $|p|$ is small,
    and we expand $f_{0}(q,p)$ into the Taylor series with respect to $p$.
      \item In $U_{2}$, $k$ is large,
    and we expand $\ave{\cos q}_{J}$ into the power series of $1/k$.
\end{itemize}
Remembering the scaling $p\simeq \Mb^{1/2}k$ in a large $k$ region,
to satisfy the above two requirements,
we have the interval of $u$ as
$ 0 < u < 1/2$. 
We note that the separatrix reaches to $p=2\Mb^{1/2}$,
but is much smaller than $\Mb^{u}$,
and hence the region $U_{1}$ includes the whole separatrix inside.
See Fig.~\ref{fig:muspace} for a schematic picture of this division.

Corresponding to the division of the $\mu$-space,
the integral $I$ is divided as
\begin{equation}
    I = I_{1} + I_{2}
\end{equation}
where
\begin{equation}
    I_{i} = \iint_{U_{i}} \ave{\cos q}_{J} f_{0}(q,p) dqdp.
\end{equation}
In the way of expansion we will neglect higher terms than $O(\Mb^{3/2})$.

\subsubsection{Region $U_{1}$}
In the region $U_{1}$, thanks to the assumption {\bf A1},
the Taylor expansion gives
\begin{equation}
    f_{0}(q,p) = f_{0}(q,0) + \dfrac{1}{2}f_{0}^{(2)}(q,0) p^{2}
    + O(p^{4}).
\end{equation}
We separately estimate contribution from the term of $O(p^{4})$
in the two subregions of $U_{1}$:
separatrix inside $U_{1,{\rm in}}$ and
separatrix outside $U_{1,{\rm out}}$.
In the separatrix inside,
the maximum of $p$ is of $O(\Mb^{1/2})$, hence we have
\begin{equation}
    \iint_{U_{1,{\rm in}}} \ave{\cos q}_{J} p^{4}dqdp
    \simeq \int_{0}^{\Mb^{1/2}} p^{4}dp
    = O(\Mb^{5/2}).
\end{equation}
This contribution is higher than $O(\Mb^{3/2})$ and is negligible.
In the separatrix outside, we have the asymptotic expansion of
$\ave{\cos q}_{J}$, \eqref{eq:asymptotic-avecos}, and we have
\begin{equation}
    \iint_{U_{1,{\rm out}}} \ave{\cos q}_{J} p^{4} dqdp
    \simeq \int_{\Mb^{1/2}}^{p_{\ast}} \dfrac{p^{4}}{k^{2}} dp
    = O(\Mb^{1+3u}).
\end{equation}
The contribution from the region $U_{1}$ is, therefore,
\begin{equation}
    \begin{split}
        	I_{1}
        	= \int_{U_{1}} 
        	\left( f_{0}(q,0) + f_{0}^{(2)}(q,0) \dfrac{p^{2}}{2} \right)
        	\ave{\cos q}_{J}dqdp
        	+ O(M^{1+3u}).
    \end{split}
\end{equation}

\subsubsection{Region $U_{2}$}
In region $U_{2}$, we use the asymptotic expansion
of $\ave{\cos q}_{J}$, \eqref{eq:asymptotic-avecos},
and $I_{2}$ is
\begin{equation}
    I_{2} = - \iint_{U_{2}} \left(
      \dfrac{1}{8k^{2}} + \dfrac{1}{16k^{4}} + O(1/k^{6})
    \right) f_{0}(q,p)dqdp.
\end{equation}
Contribution from the term of $O(1/k^{6})$ is estimated as
\begin{equation}
    \iint_{U_{2}} \dfrac{f_{0}(q,p)}{k^{6}}dqdp
    \simeq \Mb^{3} \int_{p_{\ast}}^{\infty} \dfrac{dp}{p^{6}}
    = O(\Mb^{3-5u}).
\end{equation}
Expanding the relation \eqref{eq:k-p} for small $\Mb$,
we have
\begin{equation}
    \dfrac{1}{8k^{2}} + \dfrac{1}{16k^{4}}
    = \dfrac{\Mb}{2p^{2}} + \dfrac{\Mb^{2}\cos q}{p^{4}} + O(\Mb^{3}/p^{6}),
\end{equation}
and the term of $O(\Mb^{3}/p^{6})$ gives contribution of $O(\Mb^{3-5u})$.
The integral $I_{2}$ is therefore expressed as
\begin{equation}
    \label{eq:I3-expand-1}
    I_{2} = - \iint_{U_{2}} \left(
      \dfrac{\Mb}{2p^{2}} + \dfrac{\Mb^{2}\cos q}{p^{4}}
    \right) f_{0}(q,p)dqdp + O(\Mb^{3-5u}). 
\end{equation}

We perform the integration by parts for the first term
of \eqref{eq:I3-expand-1}, and we have
\begin{equation}
    \label{eq:I3-expand-2}
    \begin{split}
        I_{2}
        & = - \int_{-\pi}^{\pi} \dfrac{\Mb}{p_{\ast}} f_{0}(q,0)dq
        - \int_{-\pi}^{\pi} \dfrac{\Mb p_{\ast}}{2} f_{0}^{(2)}(q,0)dq \\
        & - \dfrac{\Mb}{2} \iint_{U_{2}} \dfrac{f_{0}^{(1)}(q,p)}{p}dqdp
        - \iint_{U_{2}} \dfrac{\Mb^{2}\cos q}{p^{4}} f_{0}(q,p)dqdp \\
        & + O(\Mb^{3-5u}) + O(\Mb^{1+3u}),
    \end{split}
\end{equation}
where, using smallness of $p_{\ast}$, we expanded $f_{0}(q,p_{\ast})$
into the Taylor series, whose higher terms give contribution
of $O(\Mb^{1+3u})$.
We will modify the above expression to obtain the expansion
\eqref{eq:A-SC-expansion}.

Using the relation
\begin{equation}
    \int_{p_{\ast}}^{\infty} \ave{\cos q}_{J}dp
    = - \dfrac{\Mb}{2p_{\ast}}
    - \int_{p_{\ast}}^{\infty} \dfrac{\Mb^{2}\cos q}{p^{4}}dp
    + O(\Mb^{3-5u}),
\end{equation}
the sum of the first and the fourth terms of \eqref{eq:I3-expand-2} become
\begin{equation}
    \begin{split}
        & - \int_{-\pi}^{\pi} \dfrac{\Mb}{p_{\ast}} f_{0}(q,0)dq
        + \iint_{U_{2}} \dfrac{\Mb^{2}\cos q}{p^{4}} f_{0}(q,p)dqdp \\
       & = \iint_{U_{2}} f_{0}(q,0) \ave{\cos q}_{J}dqdp \\
       & \hspace*{1em} - \iint_{U_{2}} \dfrac{\Mb^{2}\cos q}{p^{4}}
        \left(f_{0}(q,p) - f_{0}(q,0) \right)dqdp.
    \end{split}
\end{equation}
Thanks to {\bf Lemma \ref{lem:f-bd}}, 
the second term is estimated as
\begin{equation}
\begin{split}
    \iint_{U_{2}} \dfrac{\Mb^{2}\cos q}{p^{4}} \left( f_{0}(q,0)  - f_{0}(q,p)\right)dqdp
    &\simeq \Mb^{2} \int_{p_{\ast}}^{\infty} \dfrac{dp}{p^{2}}\\
    &= O(\Mb^{2-u}).
    \end{split}
\end{equation}
This is higher than $O(\Mb^{3/2})$, and is negligible.

The second term is modified as
\begin{equation}
    - \int_{-\pi}^{\pi} \dfrac{\Mb p_{\ast}}{2} f_{0}^{(2)}(q,0)dq
    = - \dfrac{\Mb}{4} \iint_{U_{1}} f_{0}^{(2)}(q,0)dqdp.
\end{equation}
Modification of third term is
\begin{equation}
    \begin{split}
        & - \dfrac{\Mb}{2} \iint_{U_{2}} \dfrac{f_{0}^{(1)}(q,p)}{p}dqdp \\
        & = - \dfrac{\Mb}{2} \iint_{\mu}  
        \dfrac{f_{0}^{(1)}(q,p)}{p}dqdp
        + \dfrac{\Mb}{2} \iint_{U_{1}} f_{0}^{(2)}(q,0)dqdp \\
        & \hspace*{1em} + O(\Mb^{1+3u}),
    \end{split}
\end{equation}
where we expanded $f_{0}^{(1)}(q,p)$ in the second term of the right-hand-side
around $p=0$ with the aid of smallness of $p_{\ast}$,
and the term of $O(\Mb^{1+3u})$ comes from higher terms of the Taylor expansion.

Putting all together, we have the term $I_{2}$ as
\begin{equation}
    \begin{split}
        I_{2}
        & = \iint_{U_{2}} f_{0}(q,0) \ave{\cos q}_{J}dqdp
        - \dfrac{\Mb}{2} \iint_{\mu} \dfrac{f_{0}^{(1)}(q,p)}{p}dqdp \\
        & + \dfrac{\Mb}{4} \iint_{U_{1}} f_{0}^{(2)}(q,0)dqdp
        + O(\Mb^{3-5u}) + O(\Mb^{1+3u}).
    \end{split}
\end{equation}

\subsubsection{The whole $I$}
The above computations give the whole integral $I$ in the form
\begin{equation}
    \label{eq:I-expand-3}
    \begin{split}
        I 
        & = \iint_{\mu} f_{0}(q,0) \ave{\cos q}_{J} dqdp
        - \dfrac{\Mb}{2} \iint_{\mu} \dfrac{f_{0}^{(1)}(q,p)}{p} dqdp \\
        & + \iint_{U_{2}} f_{0}^{(2)}(q,0) 
        \left( \dfrac{p^{2}}{2}\ave{\cos q}_{J} + \dfrac{\Mb}{4} \right) dqdp \\
        & + O(\Mb^{3-5u}) + O(\Mb^{1+3u}).
    \end{split}
\end{equation}

The final modification is to extend the integral region
of the third term of Eq.~\eqref{eq:I-expand-3} to the whole $\mu$-space.
This extension can be done since contribution from the region $U_{2}$
is negligible. Indeed, using the relation \eqref{eq:k-p}
and the asymptotic expansion of $\ave{\cos q}_{J}$,
Eq.~\eqref{eq:asymptotic-avecos}, we have
\begin{equation}
    \begin{split}
        & \iint_{U_{2}} f_{0}^{(2)}(q,0)
        \left( \dfrac{p^{2}}{2}\ave{\cos q}_{J} + \dfrac{\Mb}{4} \right) dqdp \\
        & = \Mb \iint_{U_{2}} f_{0}^{(2)}(q,0) 
        \left( - \dfrac{\cos q}{8k^{2}} + O(1/k^{4}) \right) dqdp\\
        &= O(\Mb^{2-u})
    \end{split}
\end{equation}
and is higher than $O(\Mb^{3/2})$.
This extension gives the final form of $I$ as
\begin{equation}
    \label{eq:I-final}
    \begin{split}
        I 
        & = \iint_{\mu} f_{0}(q,0) \ave{\cos q}_{J} dqdp
        - \dfrac{\Mb}{2} \iint_{\mu} \dfrac{f_{0}^{(1)}(q,p)}{p} dqdp \\
        & + \iint_{\mu} f_{0}^{(2)}(q,0) 
        \left( \dfrac{p^{2}}{2}\ave{\cos q}_{J} + \dfrac{\Mb}{4} \right) dqdp \\
        & + O(\Mb^{3-5u}) + O(\Mb^{1+3u}) .
    \end{split}
\end{equation}
The optimal value of $u$ is $u=1/4$, and both
$O(\Mb^{3-5u})$ and $O(\Mb^{1+3u})$ become $O(\Mb^{7/4})$.
The above expression of the integral $I$
concludes the expansion \eqref{eq:A-SC-expansion}.

\section{Theoretical consequences}
\label{sec:theorical-consequences}
We provide some theoretical predictions obtained from the expansion
\eqref{eq:A-SC-expansion}.
For this purpose, we introduce some additional assumptions
for the initial stationary state:
\begin{enumerate}
      \item[{\bf A2}] $f_{\rm ini}$ is single-peak and spatially homogeneous,
    and is denoted by $f_{\rm ini}(p)$.
      \item[{\bf A3}] We consider a one-parameter family
    of $f_{\rm ini}$, which continuously depends on
    the parameter $\tau$.
    The family changes the stability at $\tau=\tau_{\rm c}$,
    which is called the critical point.
\end{enumerate}
For instance, the family of Maxwellians is parameterized by temperature $T$
\begin{equation}
    \label{eq:Maxwellian}
    f_{\rm ini}(p) = \dfrac{1}{2\pi\sqrt{2\pi T}} e^{-p^{2}/2T}
\end{equation}
and $T_{\rm c}=1/2$ is the critical temperature in the HMF model \cite{MA95}.
We note that the nonequilibrium phase transitions can be observed
in several families of QSSs \cite{YYY07,Antoniazzi07}.

From the above assumptions,
the functionals $L_{1/2}[f_{\rm ini}]$ and $L_{3/2}[f_{\rm ini}]$ 
can be written as
\begin{equation}
    L_{1/2}[f_{\rm ini}]
    = \Mb^{-1/2} f_{\rm ini}(0) \iint_{\mu} \ave{\cos q}_{J} dqdp =0
\end{equation}
and
\begin{equation}
    \label{eq:L32fini}
    L_{3/2}[f_{\rm ini}] 
    = \Mb^{-1/2} f_{\rm ini}^{(2)}(0) \iint \ave{\cos q}_{J}^{2} dqdp < 0.
\end{equation}
Indeed, we can show the equalities
\begin{equation}
    \label{eq:int-avecos-zero}
    \iint_{\mu} \ave{\cos q}_{J} dqdp = 0
\end{equation}
and
\begin{equation}
    \label{eq:int-p2cos-avecos2}
    \iint_{\mu} \left(
      \dfrac{p^{2}}{2}\ave{\cos q}_{J} + \dfrac{\Mb}{4} \right) dqdp
    = \Mb \iint_{\mu} \ave{\cos q}_{J}^{2} dqdp.
\end{equation}
See Appendix \ref{sec:appendix-proofs} for the proofs of these equalities.
The sign of $L_{3/2}[f_{\rm ini}]$ comes from the assumption {\bf A2},
which implies $f_{\rm ini}^{(2)}(0)<0$.

We further introduce an assumption for
the initial perturbation $g$:
\begin{enumerate}
      \item[{\bf A4}] $g$ has no Fourier zero mode with respect to $q$.
\end{enumerate}
With this assumption $g$ satisfies the normalization condition
\eqref{eq:intg0}.
The assumption {\bf A4} eliminates $L_{1}[g]$.

Omitting the higher order terms,
we have the self-consistent equation of the form
\begin{equation}
    \label{eq:A-SC-homo}
    \epsilon L_{1/2}^{1} (M+h)^{1/2}
    - D M +(1-D)h 
    + L_{3/2}^{0}(M+h)^{3/2} =0,
\end{equation}
where $L_{1/2}^{1}=L_{1/2}[g], L_{3/2}^{0}=L_{3/2}[f_{\rm ini}]$, and
\begin{equation}
    D = 1 - L_{1}[f_{\rm ini}]
    = 1 + \dfrac{1}{2} \iint_{\mu} \dfrac{f_{\rm ini}^{(1)}(p)}{p} dqdp.
\end{equation}
The functional $D$ represents the stability functional
for a single-peak homogeneous even distribution \cite{YYY04},
and positive (resp. negative) $D$ implies that $f_{\rm ini}$ is stable
(resp. unstable).
For instance, one can directly confirm
\begin{equation}
    \label{eq:D}
    D=(T-T_{\rm c})/T
\end{equation}
for the Maxwellians \eqref{eq:Maxwellian}.
The functional $D$ is small around the critical point,
and hence we keep the last term of $O(\Mb^{3/2})$.

\begin{figure}[tb]
	\centering
	\includegraphics[width=7cm]{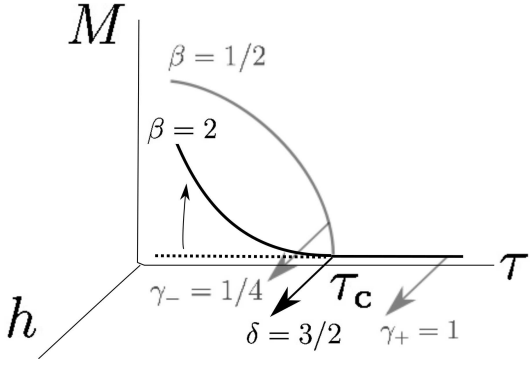}\\
	\caption{Schematic picture of QSS families and critical exponents.
          Bold lines represents stable QSSs,
          and dashed unstable, which goes to the bold line
          indicated by $\beta=2$.
          The straight line on $M=0$, which is slightly shifted
          for a graphical reason,
          corresponds to the family of $f_{\rm ini}$
          mentioned in the assumption {\bf A3}.
          We assumed that $f_{\rm ini}$ is stable (resp. unstable)
          for $\tau>\tau_{\rm c}$ (resp. $\tau<\tau_{\rm c}$)
          without loss of generality.
          The exponents $\gamma_{+}$ and $\gamma_{-}$ are
          obtained in Refs. \cite{AP12,SO12} and \cite{SO13}
          respectively.
          The exponent $\beta$ for prepared stable inhomogeneous states 
          is $1/2$ in usual case, as we will explain in Sec.~\ref{sec:Scaling}. 
	 }
	\label{fig:schematic}
\end{figure}

Considered QSS families and the critical exponents are summarized
as a schematic picture in Fig.~\ref{fig:schematic}.
In the following Secs.~\ref{sec:theory-response-h}
and \ref{sec:theory-response-ep},
we will show the critical exponents $\delta=3/2$ and $\beta=2$
with quantitative predictions of $M$.

\subsection{Response to external field}
\label{sec:theory-response-h}
We set $\epsilon=0$ and $h>0$, and observe response to the external field.
The self-consistent equation is reduced to
\begin{equation}
    \label{eq:A-SC-epsilon0}
    - D M +(1-D)h
    + L_{3/2}^{0}(M+h)^{3/2} =0.
\end{equation}

\subsubsection{Stable off critical}
We first consider the off-critical situation
with $f_{\rm ini}$ stable, $D>0$. 
Solving Eq.~\eqref{eq:A-SC-epsilon0} recursively 
and taking into account the truncated 
term of order $O((M+h)^{7/4})$, we have
\begin{equation}
    \label{eq:M-response-h}
    M = \chi h + \dfrac{L_{3/2}^{0}}{D}(\chi+1)^{3/2} h^{3/2} + O(h^{7/4}),
\end{equation}
where
\begin{equation}
    \chi = \dfrac{1-D}{D}.
\end{equation}
The leading order is identical to the linear response
of homogeneous states \cite{AP12,SO12}.

One remarkable difference between the present method
and the linear response theory is
that the latter is based on approximated solutions 
to the Vlasov equation constructed by perturbation technique.
Higher order computations of the perturbation procedure
give vanishing $O(h^{2})$ term due to symmetry,
and the nonvanishing next leading order is of $O(h^{3})$,
while the present method gives the next leading of $O(h^{3/2})$.
The two methods coincide in the linear regime,
but do not in the nonlinear regime.
In Sec.~\ref{sec:numerics},
we will numerically confirm that the present method successfully predicts
values of the magnetization even in a large $h$ regime.

\subsubsection{On critical}
\label{sec:reponse-h-oncritical}

In addition to the nonlinear response in off-critical situation,
the present method has another advantage against the linear response theory,
which is the prediction at the critical point.
The linear response theory can not apply to the critical point,
since the susceptibility $\chi$ diverges.
However, the present method gives the critical exponent $\delta$
defined in Eq.~\eqref{eq:critical-exponent-delta}.
Let us set the parameter $\tau$ as the critical value $\tau_{\rm c}$,
and $D=0$ accordingly. The solution to the self-consistent equation
\eqref{eq:A-SC-epsilon0} is
\begin{equation}
    \label{eq:exponent-delta}
    M = \left( \dfrac{h}{-L_{3/2}^{0}} \right)^{2/3} - h.
\end{equation}
Note that $L_{3/2}^{0}<0$ and hence the first term of the right-hand-side
is meaningful.
The right-hand-side is dominated by $h^{2/3}$ for small $h$,
and hence the critical exponent $\delta$ for an isolated system 
is evaluated as $\delta=3/2$.
It is worth noting that the classical mean-field theory gives
the isothermal critical exponent $\delta=3$.

\subsubsection{Scaling relation}
\label{sec:Scaling}

Apart from the initial stationary homogeneous state $f_{\rm ini}(p)$,
we consider the family of QSSs parameterized by $\tau$.
We consider the Jeans type family
\begin{equation}
    \label{eq:Jeans}
    f(q,p;\tau) = F(p^{2}/2-M(\tau)\cos q;\tau)
\end{equation}
with $F$ smooth, and assume that $f$ depends on $M$
solely through the Hamiltonian.
We set that states of the family are stable homogeneous
for $\tau>\tau_{\rm c}$, and stable inhomogeneous for $\tau<\tau_{\rm c}$,
without loss of generality.
For such a family we can show the scaling relation
\begin{equation}
    \label{eq:scaling-betagamma}
    \gamma_{-} = \beta/2
\end{equation}
from the linear response theory \cite{SO13},
where the critical exponents $\beta$ and $\gamma_{-}$ are defined
in the inhomogeneous side by
\begin{equation}
    M \propto (\tau_{\rm c}-\tau)^{\beta}, \quad
    \chi = \left. \dfrac{{\rm d}M}{{\rm d}h}
    \right|_{h\to 0} \propto (\tau_{\rm c}-\tau)^{-\gamma_{-}}
\end{equation}
respectively.
The value of $\beta$ is $1/2$ in general,
since the family \eqref{eq:Jeans} must satisfy
the self-consistent equation for $M$,
and it is expanded around $M=0$ as the Landau's phenomenological theory:
\begin{equation}
	D_{\rm homo} M + BM^{3} + O(M^{5}) = 0, 
\end{equation}
where the stability functional $D_{\rm homo}$ is evaluated for
the homogeneous state $F(p^{2}/2;\tau)$ and is positive (resp. negative)
in the homogeneous side $\tau>\tau_{\rm c}$
(resp. the inhomogeneous side $\tau<\tau_{\rm c}$)
since the homogeneous state is stable (resp. unstable).
We have assumed $B>0$.

The scaling relation \eqref{eq:scaling-betagamma} ensures
the scaling relation
\begin{equation}
	\label{eq:scaling-relation}
	\gamma_{-} = \beta (\delta-1)
\end{equation}
with the obtained exponent $\delta=3/2$.
The scaling relation \eqref{eq:scaling-relation} is, therefore,
extended from the family of thermal equilibrium states in the isothermal system
to families of QSSs in the isolated system.
The critical exponents are summarized in Table \ref{tab:critical-exp}.

\begin{table}[hb]
    \caption{Critical exponents for asymptotic states in
      isothermal and isolated systems.
      The value of $\beta$ in the isolated system
      depends on choice of a family of QSSs,
      but $\beta=1/2$ in general.
      Moreover, the relation $\gamma_{-}=\beta/2$ holds.
      These exponents satisfy the scaling relation
      $\gamma_{-} = \beta(\delta - 1)$ in both systems.}
    \label{tab:critical-exp}
    \begin{tabular}[b]{c|c|c}
        \hline
        \hline
        Critical Exp. & Isothermal & Isolated QSSs \\
        \hline
        $\beta$ &1/2 & 1/2  \\
        $\gamma_{-}$ & 1 & 1/4  \\
        $\delta$ & 3 & 3/2  \\
        \hline
        \hline
        \end{tabular}
\end{table}

\subsection{Response to perturbation}
\label{sec:theory-response-ep}
We set $h=0$ and $\epsilon>0$, and observe response to perturbation.
The self-consistent equation is reduced to
\begin{equation}
    \label{eq:A-SC-h0}
    \epsilon L_{1/2}^{1} M^{1/2}
    - D M
    + L_{3/2}^{0} M^{3/2} = 0.
\end{equation}

Taking the limit $\epsilon\to 0$, we have
\begin{equation}
    M = \left\{
      \begin{array}{ll}
          0 & \quad\text{($f_{\rm ini}$ stable)} \\
          0,~ \left(\dfrac{D}{L_{3/2}^{0}}\right)^{2} & \quad\text{($f_{\rm ini}$ unstable).}
      \end{array}
    \right.
\end{equation}
As the Maxwellians, we may expect that $D$ for $f_{\rm ini}$
linearly depends on the parameter $\tau-\tau_{\rm c}$ in general. 
The asymptotic magnetization for unstable $f_{\rm ini}$, therefore,
has the scaling
\begin{equation}
    \label{eq:Response-ep-scaling}
    M\propto (\tau-\tau_{\rm c})^{2},
\end{equation}
which is consistent with a theoretical analysis for plasma system \cite{JDC95}
and with numerics for self-gravitating system \cite{AVI05}.
The scaling implies that the critical exponent $\beta$
for this family is $\beta=2$.

Coming back to nonzero $\epsilon$ but at the critical point,
we have the response
\begin{equation}
    \label{eq:Response-ep-critical}
    M = 0,~ \dfrac{L_{1/2}^{1}}{-L_{3/2}^{0}}~ \epsilon. 
\end{equation}
Considering that the linear response at critical point diverges,
for positive $L_{1/2}^{1}$,
we may expect that the $M=0$ branch is unstable
and the other one is stable.
This expectation will be confirmed numerically in Sec.~\ref{sec:numerics}.

For stable homogeneous $f_{\rm ini}$,
thanks to $L_{1/2}^{1}>0$ and $L_{3/2}^{0}<0$, 
we have the non-zero solution to the self-consistent equation
\eqref{eq:A-SC-h0} as
\begin{equation}
    \label{eq:M-response-epsilon}
    M
    = \left( \dfrac{D-\sqrt{D^{2}-4\epsilon L_{1/2}^{1}L_{3/2}^{0}}}{2L_{3/2}^{0}}
    \right)^{2}.
\end{equation}

\section{Numerical tests}
\label{sec:numerics}

We perform numerical computations of the Vlasov
equation by using the semi-Lagrangian method \cite{PdB10} 
with the time step $\Delta t = 0.05$. 
The $\mu$ space $(-\pi,\pi]\times \mathbb{R}$ is truncated as 
$(-\pi, \pi] \times [-4, 4]$, and each axis is divided into $N$ bins. 
We call $N$ the grid size. 
The initial condition is set as the Maxwellian and small perturbation:
\begin{equation}
    \label{eq:fini-g}
    f_{\rm ini}(p) = \frac{1}{2\pi\sqrt{2\pi T}} e^{-p^2/2T},
    \quad
    g(q,p) = f_{\rm ini}(p) \cos q.
\end{equation}
The initial stationary state $f_{\rm ini}$ is stable for $T>T_{\rm c}=1/2$,
and is unstable for $T<T_{\rm c}$,
and this family satisfies the assumptions {\bf A2} and {\bf A3}. 
The above symmetric perturbation $g$ is suitable
for the considering situation {\bf H4},
satisfies {\bf H3},
and satisfies the assumptions {\bf A1} and {\bf A4}.

For the initial state \eqref{eq:fini-g},
we have the values of $L_{1/2}^{1}$ and $L_{3/2}^{0}$ as
\begin{equation}
    \label{eq:numerics-coefficients}
    L_{1/2}^{1} = \xi f_{\rm ini}(0)
    = \dfrac{\xi}{2\pi\sqrt{2\pi T}}, \quad
    L_{3/2}^{0} = \xi f_{\rm ini}^{(2)}(0)
    = - \dfrac{L_{1/2}^{1}}{T},
\end{equation}
where the factor $\xi$ is defined and computed as
\begin{equation}
    \begin{split}
        \xi & = M^{-1/2} \iint \ave{\cos q}^{2}_{J} dqdp
        \simeq 5.17.
    \end{split}
\end{equation}
The value of functional $D$ is obtained as Eq.~\eqref{eq:D}.

We compute the asymptotic value $M$ of the order parameter
as the time average of $\mathcal{M}_{x}[f]$,
defined by Eq.~\eqref{eq:Mx}.

\subsection{Response to external field}
\begin{figure}[tb]
	\centering
	\includegraphics[width=8cm]{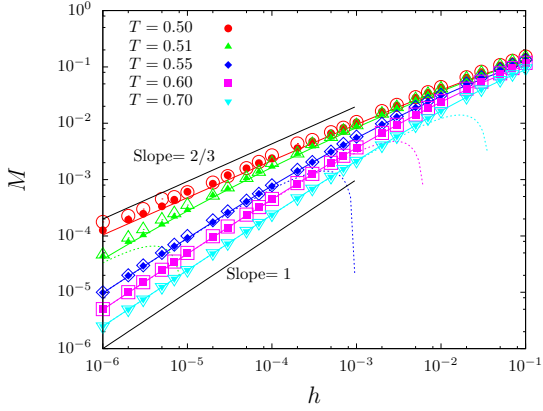}\\
	\caption{(color online) 
          External field $h$ dependence of response. 
          $T=0.50(=T_{\rm c})$ (red circles), $0.51$(green triangles), 
          $0.55$ (blue diamonds), $0.60$ (purple squares) 
          and $0.70$ (light blue inverse triangles).
          Grid sizes are $N=256$ (open symbols) and 
          $N=512$ (filled symbols). 
          The value of $M$ is computed taking time averages
          over the time interval $[500, 3000]$.  
          The solid curves represent solutions to the self-consistent
          equation \eqref{eq:A-SC-epsilon0} for each temperature $T$,
          and $T$ increases from top to bottom.
          The dashed curves are from \eqref{eq:M-response-h}.
          Two black solid lines with slopes $2/3$ and $1$ are guides of eyes.
	}
	\label{fig:Response_h}
\end{figure}

We set $\epsilon=0$ and $h>0$ with $T\geq T_{\rm c}$,
and observe $M$ as functions of $h$ in Fig.~\ref{fig:Response_h}.
We used the two grid sizes of $N=256$ and $512$,
and both sizes are in good agreements with each other.
Thus, these grid sizes are large enough for observing responses.

We stress the following two observations:
($h$-i) At the critical temperature $T=T_{\rm c}$,
the slope $2/3$ corresponding to the critical exponent $\delta=3/2$
is successfully observed, and the slope goes to $1$ with increasing $T$
as predicted by the linear response theory \cite{AP12, SO12}.
We note that the slope seems to change smoothly
between $2/3$ and $1$,
but it must be $1$ in the limit of small $h$
except for $T=T_{\rm c}$.
($h$-ii) The solutions to the self-consistent equation \eqref{eq:A-SC-epsilon0},
solid curves in Fig.~\ref{fig:Response_h},
are in good agreement with numerical simulations
even in a large $h$ regime, beyond the linear response regime.

The recursive solution \eqref{eq:M-response-h},
dashed curves in Fig.~\ref{fig:Response_h},
does not provide good predictions for $T$ close to $T_{\rm c}$,
since the omitted part as higher order terms includes $\chi$
which becomes large as $T$ approaches to $T_{\rm c}$.

\subsection{Response to perturbation}
We numerically examine the three theoretical consequences:
($\epsilon$-i) $M$ linearly depends on $\epsilon$
at the critical point, \eqref{eq:Response-ep-critical}.
($\epsilon$-ii) Nonlinear response of $M$ for $f_{\rm ini}$ stable case
with the expression \eqref{eq:M-response-epsilon}.
($\epsilon$-iii) The scaling \eqref{eq:Response-ep-scaling}
for $f_{\rm ini}$ unstable case.

\begin{figure}[tb]
	\centering
	\includegraphics[width=8cm]{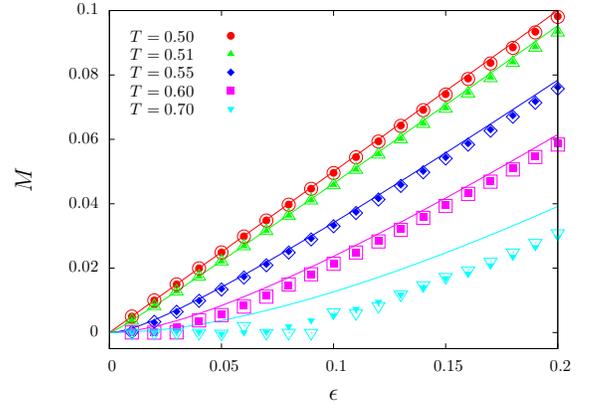}\\
	\caption{
	(color online) 
	Perturbation $\epsilon$ dependence of response with several $T$ 
	in the disordered phase ($T\geq T_{\rm c}=1/2$).
	Lines are drawn by \eqref{eq:M-response-epsilon},
        and points are by numerical simulations.
        Open symbols are for the grid size $N=256$,
        and filled are for $N=512$.
	Temperature $T$ is $T = 0.50(=T_{\rm c})$ (red circles), 
	$0.51$ (green triangles), $0.55$ (blue diamonds),
        $0.60$ (purple squares), and $0.70$ (light blue inverse triangles). 
        Time averages are taken in the interval $t\in [500,3000]$.
	}
	\label{fig:Response_ep_HighT}
\end{figure}

Numerical tests for the consequences ($\epsilon$-i) and ($\epsilon$-ii)
are exhibited in Fig.~\ref{fig:Response_ep_HighT}.
The coefficients \eqref{eq:numerics-coefficients} gives
the theoretical response at the critical temperature as
\begin{equation}
    M = T_{\rm c} \epsilon, \qquad T_{\rm c}=1/2,
\end{equation}
and the numerical response perfectly coincides with this theoretical
prediction.
Apart from the critical temperature,
the numerical responses are in good agreements with the theory
for temperature close to the critical point.
As $T$ increases, the agreement becomes worse quantitatively,
but is still good qualitatively
except for a threshold like dependence on $\epsilon$,
for instance $M=0$ below the threshold $\epsilon_{\rm th}\simeq 0.07$
for $T=0.7$,
cannot be reproduced by the present theory.

\begin{figure}[tb]
	\centering
	\includegraphics[width=8cm]{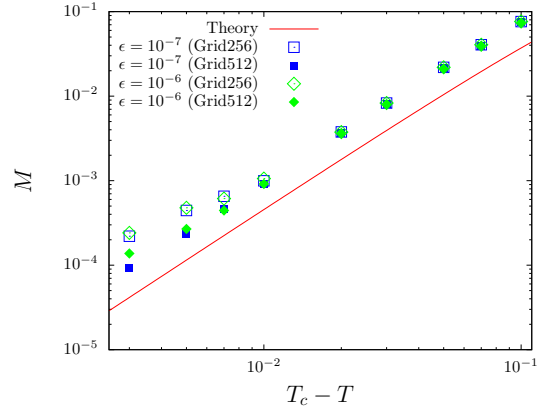}\\
	\caption{
	(color online) 
	Asymptotic value of $M$ with small perturbation $\epsilon$ 
	for the unstable Maxwellian initial states.
	The solid red line is the theoretical prediction,
        and points are numerically computed values.
        $\epsilon=10^{-7}$ (blue squares)
        and $10^{-6}$ (green diamonds).
        The grid sizes are $N=256$ (open symbols)
        and $N=512$ (filled symbols).
	Time average interval is $[1000, 3000]$. 
	}
	\label{fig:Response_ep_LowT}
\end{figure}

The scaling for the unstable $f_{\rm ini}$, ($\epsilon$-iii),
is confirmed in Fig.~\ref{fig:Response_ep_LowT}.
With the coefficients \eqref{eq:numerics-coefficients}
we have the theoretical scaling as
\begin{equation}
    M = \dfrac{(T-T_{\rm c})^{2}}{(L_{1/2}^{1})^{2}}
    = \dfrac{(2\pi)^{3}T}{\xi^{2}} (T-T_{\rm c})^{2}.
\end{equation}
The absolute values of numerical responses
are slightly larger than the theoretical ones,
but the scaling is perfect.

\begin{figure}[t]
	\centering
	\includegraphics[width=8cm]{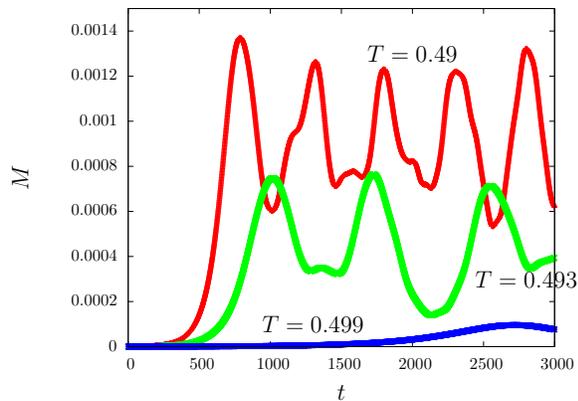}\\
	\caption{
	(color online) 
        Temporal evolutions of $M$ for unstable $f_{\rm ini}$ case
        with $\epsilon=10^{-6}$ and $h=0$.
        $T=0.49, 0.493$ and $0.499$ from top to bottom.
	}
	\label{fig:oscillation}
\end{figure}

We remark that it is not easy to numerically check
validity of the scaling relation \eqref{eq:scaling-relation}
for the family with $\beta=2$ starting from the unstable
homogeneous $f_{\rm ini}$.
For getting the exponent $\gamma_{-}$,
we have to compute the asymptotic state accurately,
but the state oscillates in the computing time as shown
in Fig.~\ref{fig:oscillation}.
Studying validity of the hypothesis {\bf H0}
and the scaling relation for this family
is left as a future work.

\subsection{On discrepancies between theory and numerics}
\label{sec:discrepancy}
We observed quantitative discrepancies between the theory and numerics
in the response for perturbation.
In high-temperature side, the agreement becomes worse
as temperature goes up. In low-temperature side,
the numerical response is systematically larger than
the theoretical prediction.
For each discrepancy,
we propose a possible explanation:
The Landau damping cannot be neglected for the former,
and the transient field is not small for the latter. 
We will discuss that 
the two explanations come from breaks of the hypotheses
{\bf H2} and {\bf H1} respectively.

We remark that the hypotheses and the assumptions are satisfied
in the numerical setting except for {\bf H0}, {\bf H1}, {\bf H2}
and {\bf A0}, and we find numerically that {\bf A0} is satisfied.
The hypothesis {\bf H0} breaks for the latter case
as shown in Fig.~\ref{fig:oscillation},
where the oscillation does not tend to vanish in the computing time scale.
However, the breaking may not be serious
for predicting the value of asymptotic magnetization $M$,
since effects of the oscillation could be suppressed by taking time averages,
and the value of magnetization could be approximately obtained.
We, therefore, focus on the hypotheses {\bf H1} and {\bf H2}.

The stability functional $D$ is zero at the critical point,
and hence the Landau damping rate is zero,
since $D$ is obtained by setting the frequency zero in the dispersion function. 
In this case, the nonlinear trapping effect dominates
the Landau damping as mentioned in {\bf Remark \ref{remark}},
and the theory and numerics are in good agreement.

Increasing temperature, the Landau damping rate becomes 
larger, so that we cannot neglect it. 
From the view point of physics, 
this is interpreted as follows: 
The large Landau damping rate 
results in that 
the nonlinear trapping becomes harder,
since the trapping requires the condition
that the Landau damping time scale is much longer
than the trapping time scale. 

\begin{figure}[tb]
	\centering
	\includegraphics[width=4.2cm,height=4cm]{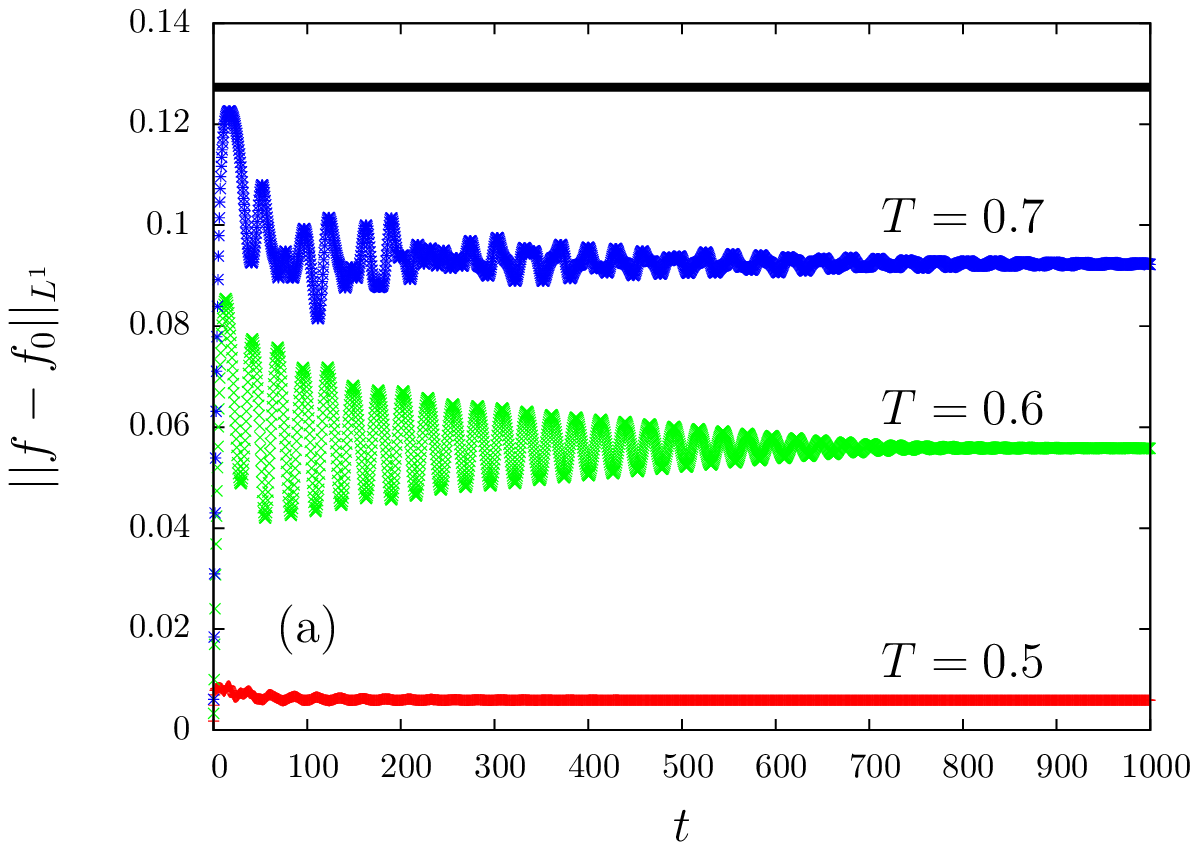}
	\includegraphics[width=4.2cm,height=4cm]{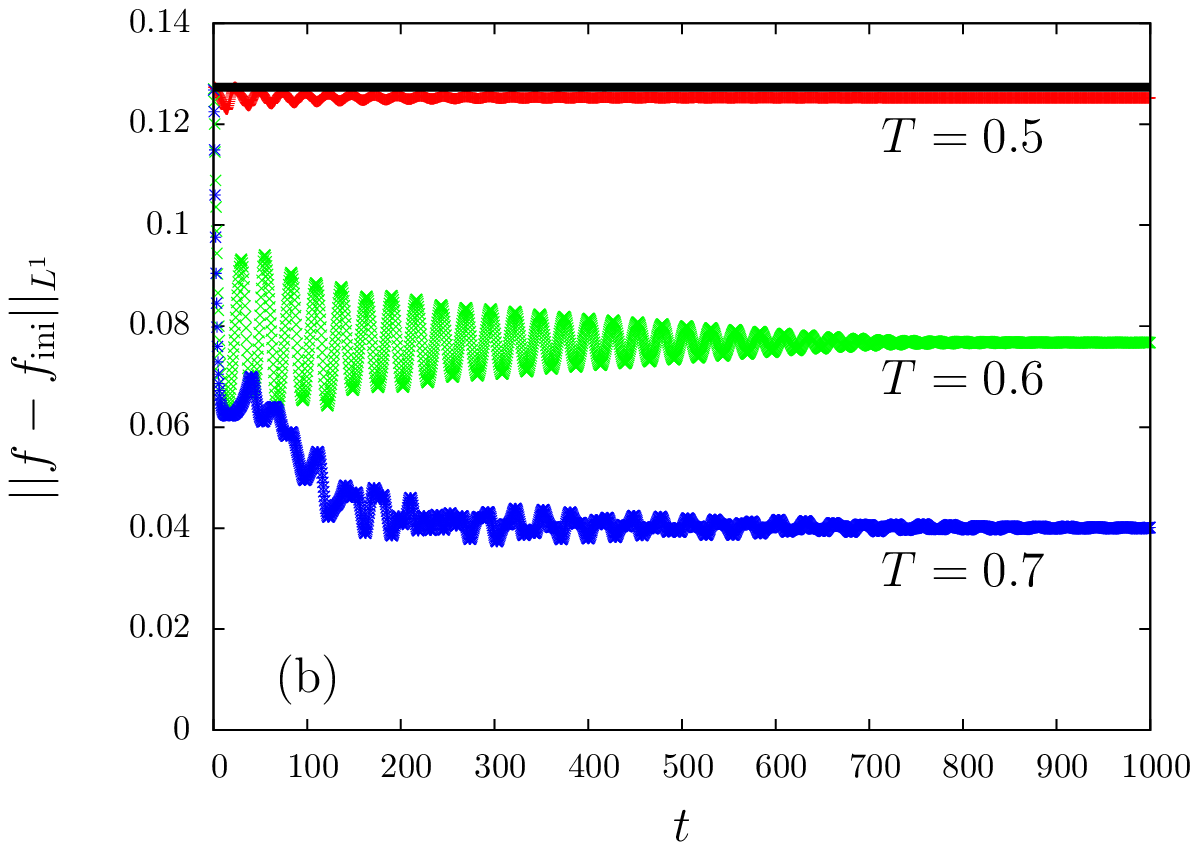}
	\caption{
	(color online) 
        Temporal evolutions of $L^{1}$-norm of
        (a) $f-f_{0}$ and (b) $f-f_{\rm ini}$.
        The grid size is $N=512$.
        The perturbation level is $\epsilon=0.2$.
        The black straight line at the level $2\epsilon/\pi$
        represents $L^{1}$-norm of $f_{\rm ini}-f_{0}=\epsilon g$,
        which does not depend on temperature.
	}
	\label{fig:norm}
\end{figure}

The above picture is supported by observing the $L^{1}$-norms
of $f-f_{0}$ and $f-f_{\rm ini}$ in Fig.~\ref{fig:norm}.
From Fig.~\ref{fig:norm}(a) reporting $||f-f_{0}||_{L^{1}}$,
the asymptotic states are trapped at closer states to
the initial states $f_{0}$ as $T$ approaches to the critical point,
$T=T_{\rm c}=1/2$.
On the other hand, from Fig.~\ref{fig:norm}(b)
reporting $||f-f_{\rm ini}||_{L^{1}}$,
the initial perturbation $g$ strongly damps as $T$ increases.
The $L^{2}$-norms give the same tendency,
though they are not reported.
We further observe from Fig.~\ref{fig:norm}
that $f$ stays close to $f_{\rm ini}$ and $f_{0}$,
and the hypothesis {\bf H1} is satisfied.
As a result, we may conclude that the discrepancy
in the high-temperature side comes from breaking {\bf H2}.

In low-temperature side, the initial homogeneous Maxwellians
are unstable, and the asymptotic states are not necessary
to be close to the initial states.
However, the present theory assumes that the transient field
is small in the decomposition \eqref{eq:fgt},
and this assumption induces that the asymptotic state
must be close to the initial state.
We may therefore conclude that the hypothesis {\bf H1}
breaks for the unstable Maxwellians $f_{\rm ini}$.

\section{Conclusion and remarks}
\label{sec:conclusion}

We have investigated the asymptotic states
of stationary initial distributions with small
external field and/or perturbation.
The method of asymptotic-transient decomposition and T-linearization
is applied to the HMF model, which is a simple toy model
of a ferromagnetic body. The theory is examined
by numerical simulations of the Vlasov equation
for the initial distributions of the Maxwellians.

The present theory unifies the two known theories:
the linear response theory \cite{AP12, SO12}
for the response to the external field,
and the bifurcation theory \cite{JDC95}
with constructing the unstable manifold of an unstable homogeneous state.
We emphasize that the present theory has further advantages
beyond the unification of the known ones: The theory
(i) captures the nonlinear response both to the external field
and to perturbations, and (ii) is applicable on the critical point.
For the latter advantage, we reported that 
the magnetization linearly depends on strength of perturbation,
and that the critical exponent $\delta$,
defined by $M\propto h^{1/\delta}$ at the critical point,
is the strange value of $\delta=3/2$, while the classical mean-field theory
gives $\delta=3$.
Interestingly, this critical exponent satisfies the scaling relation
$\gamma_{-}=\beta(\delta-1)$
with the aid of another scaling relation $\gamma_{-}=\beta/2$ \cite{SO13}. 
We stress that the above critical exponent $\delta$ and the scaling relations
are derived without assuming the thermal equilibrium states
as the initial stationary states.
The exponent and the scaling relations are, therefore, true
not only for the thermal equilibrium states
but also for Jeans type QSSs 
in the isolated long-range system. 

We have constructed the framework of the nonlinear response,
which is, for instance, the self-consistent equation \eqref{eq:A-SC}
in the HMF model. 
However, we have several remaining works: 
We have assumed that the order parameter $M$ is small
to expand the self-consistent equation into the power series of $M$.
One of the remaining works is to analyze 
the self-consistent equation for inhomogeneous initial stationary states
for reproducing the linear response theory \cite{SO12} for instance. 
Providing the asymptotically self-consistent
equation for any perturbations, beyond the hypothesis {\bf H4},
is also remained.
Another remaining work is improvement of the discrepancy
in the high temperature region discussed in Sec.~\ref{sec:discrepancy}
including the threshold like dependence on $\epsilon$.
We discussed the competition between the Landau damping
and the nonlinear trapping effect,
and hence one possible modification
is to include the Landau damping into the present theory.
Universality of the scaling relations \eqref{eq:scaling-relation}
for a wide class of systems is an interesting problem.
We focused on magnetization in this paper,
but studying the asymptotic distribution function $f_{\rm A}$ itself
is also an open problem, since $f_{\rm A}$ may have a cusp at the separatrix
as the function $\ave{\cos q}_{J}$ shown in Fig.~\ref{fig:cosq_J},
and the cusp might be unphysical.

We remark that the exponential dampings are also derived
for the 2D Euler system \cite{RJB70},
and the creation of small traveling clusters 
by the nonlinear effects is also discussed \cite{SO14}
along the same strategy with the Vlasov case.
Thus, one may expect to construct a similar nonlinear response theory
for the Euler system, which is obeyed by a similar equation
with the Vlasov equation.

We end this article with remarking on another study
of the nonlinear dynamics in the HMF model with the external field.
Pakter and Levin \cite{RP13} derived
equations of temporal evolution for macro variables,
and observed oscillations of variables.
Such oscillating phenomena are out of scope from the present theory,
since the theory assumes asymptotically stationary states.

\acknowledgments
We are grateful to J. Barr\'e for valuable discussions on this topic. 
SO is supported by the JSPS Research Fellowships for Young Scientists (Grant No. 254728). 
YYY acknowledges the support of a Grant-in-Aid for Scientific Research (C) 23560069.

\appendix

\section{Derivation of Eq.~\eqref{eq:fA}}
\label{app:B0}
We have used the fact \eqref{eq:fA} that the asymptotic part $f_{\rm A}$ can be picked up from $f$ by 
use of the Bohr transform $\mathcal{B}_0[f]$. 
Let us show this statement:

\begin{lemma}
Let $a(t)$ be a bounded function having the limit
$\displaystyle{\lim_{t\to\infty}a(t)}$.
Then the limit is expressed by the Bohr transform of $a(t)$ as
\begin{equation}
	\label{eq:Frechet}
        \lim_{t\to\infty} a(t) = \mathcal{B}_0[a].
\end{equation}
\end{lemma}
{\bf Proof: } 
We denote the limit as
$\displaystyle \lim_{t \to \infty} a(t)=a_{\infty}$,
and show the equality $a_{\infty}=\mathcal{B}_{0}[a]$.
From the assumption of convergence,
for any $\epsilon > 0$, there exists $T > 0$ such that 
\begin{equation}
	\label{eq:epsilon1}
	\left| a(t) - a_{\infty}\right| < \epsilon, \quad \forall t > T. 
\end{equation}
Then, the integral in the definition of $\mathcal{B}_0$ is evaluated for $\sigma > T$ as follows: 
\begin{equation}
	\begin{split}
		&\left|\frac{1}{\sigma}\int_0^\sigma a(t)~dt - a_{\infty}\right| 
		= 
		\left|\frac{1}{\sigma}\int_0^\sigma \left(a(t)- a_{\infty}\right)~dt \right| \\
		&\leq 
		\frac{1}{\sigma}\int_0^\sigma \left|a(t)- a_{\infty}\right|~dt\\
		&= 
		\frac{1}{\sigma}\int_0^T \left|a(t)- a_{\infty}\right|~dt 
		+ \frac{1}{\sigma}\int_T^\sigma \left|a(t)- a_{\infty}\right|~dt \\
		&\leq 
		\frac{2T \sup_{t\geq0}|a(t)|}{\sigma} + \frac{\left(\sigma-T\right)\epsilon}{\sigma} 
		\to \epsilon \quad (\sigma \to \infty). 
	\end{split}
\end{equation}
Since the positive $\epsilon$ is chosen arbitrarily, 
we have Eq. \eqref{eq:Frechet}. $\blacksquare$

\section{Elliptic integrals}
\label{app:elliptic}
The Legendre's elliptic integrals of the first and the second kinds
are defined as
\begin{equation}
	\label{eq:Legendre}
	\begin{split}
		F(\phi, k) &\equiv \int_0^\phi \frac{d\varphi}{\sqrt{1-k^2\sin^2 \varphi}}, \\
		E(\phi, k) & \equiv \int_0^\phi \sqrt{1-k^2\sin^2 \varphi} d\varphi, 
	\end{split}
\end{equation}
respectively \cite{ETW27}. 
The complete elliptic integrals of the first and the second kinds
are defined by taking $\phi=\pi/2$ as
\begin{equation}
    K(k) = F(\pi/2, k), \quad E(k) = E(\pi/2,k)
\end{equation}
respectively. The Jacobian elliptic function $\sn(z,k)$ is defined as
\cite{MA72}
\begin{equation}
    \sn\left(F(\phi, k),k\right) = \sin \phi.
\end{equation}

With the above preparations, we can compute $\ave{\cos q}_{J}$ as follows.
In action angle variables, $\cos q$ is expressed as
\begin{equation}
    \cos q = \left\{
      \begin{array}{ll}
          1 - 2k^{2}{\rm sn}^{2}\left( \dfrac{2K(k)}{\pi}\theta,k\right), & k<1, \\
          1 - 2{\rm sn}^{2}\left( \dfrac{K(1/k)}{\pi}\theta, \dfrac{1}{k} \right), & k>1.
      \end{array}
    \right.
\end{equation}
For $k<1$, inside separatrix, the average is
 \begin{equation}
     \label{eq:avecos-inside}
     \begin{split}
         \ave{\cos q}_{J}
         & = \dfrac{2}{\pi} \int_{0}^{\pi/2} \left[
           1 - 2k^{2}{\rm sn}^{2}\left( \dfrac{2K(k)}{\pi}\theta,k\right)
         \right] d\theta \\
         & = \dfrac{1}{K(k)} \int_{0}^{\pi/2}
         \dfrac{1-2k^{2}\sin^{2}\phi}{\sqrt{1-k^{2}\sin^{2}\phi}} d\phi \\
         & = \dfrac{2E(k)-K(k)}{K(k)}, \quad (k<1)
     \end{split}
 \end{equation}
where we used the change of variable
\begin{equation}
    \theta = \dfrac{\pi}{2K(k)} F(\phi,k).
\end{equation}
Similarly, we can compute the average $\ave{\cos q}_{J}$ for $k>1$,
outside separatrix.

The complete elliptic integrals are expanded into the Taylor series as
\begin{equation}
    \begin{split}
        K(k)
        & = \dfrac{\pi}{2} \sum_{n=0}^{\infty}
        \left( \dfrac{(2n-1)!!}{(2n)!!} \right)^{2} k^{2n} \\
        & = \dfrac{\pi}{2} \left(
          1 + \dfrac{k^{2}}{4} + \dfrac{9}{64} k^{4} + \dfrac{25}{256} k^{6}
          + \dfrac{35^{2}}{128^{2}} k^{8}
          + \cdots \right) 
    \end{split}
\end{equation}
and
\begin{equation}
    \begin{split}
        E(k)
        & = \dfrac{\pi}{2} \sum_{n=0}^{\infty}
        \left( \dfrac{(2n-1)!!}{(2n)!!} \right)^{2} \dfrac{k^{2n}}{1-2n} \\
        & = \dfrac{\pi}{2} \left(
          1 - \dfrac{k^{2}}{4} - \dfrac{3}{64} k^{4} - \dfrac{5}{256} k^{6}
          - \dfrac{175}{128^{2}} k^{8}
          + \cdots \right).
    \end{split}
\end{equation}
The function $\ave{\cos q}_{J}$ is therefore asymptotically expanded as
\begin{equation}
    \label{eq:asymptotic-avecos}
    \ave{\cos q}_{J}
    = - \dfrac{1}{8k^{2}} - \dfrac{1}{16k^{4}} + O(k^{-6})
\end{equation}
for $k\gg 1$.

\section{Proof of Lemma \ref{lem:f-bd}}
\label{app:lemma2}
Let us show Lemma \ref{lem:f-bd}. \\
{\bf Proof:}
Let $\delta$ be a positive number satisfying $\delta < 1$ 
and let us recall $f(q,p)$ satisfies the hypothesis {\bf H3}, 
that is, $f^{(1)}(q,0) = 0$. 
Then, the Taylor's theorem says that, for each $p \in(-\delta, \delta)$
there exists $\xi \in (0, \delta)$ such that 
\begin{equation}
	f(q,p) - f(q,0) = \frac{f^{(2)}(q,\xi)}{2} p^2, 
\end{equation}
where the right-hand-side one is called Lagrange form of the remainder \cite{ETW27}. 
We hence obtain the inequality 
\begin{equation}
	|f(q,p) - f(q,0)| 
	< c_1 p^2, \quad \forall p \in (-\delta, \delta),
\end{equation}
where $c_1$ is a positive constant satisfying 
\begin{equation}
	\begin{split}
		c_1 &>  \sup_{(q,\xi) \in (-\pi,\pi]\times[0, \delta] }\frac{|f^{(2)}(q, \xi)|}{2} .
	\end{split}
\end{equation}
For $p$ satisfying $|p| \geq \delta$, the inequality 
\begin{equation}
	|f(q,p) - f(q,0)| < c_2, 
\end{equation}
holds for some $c_2 > 0$. 
The positive constant $c_2$ can be 
chosen so that 
\begin{equation}
	\label{eq:c2}
	c_2 = 2 \sup_{(q,p) \in (-\pi, \pi] \times \mathbb{R}} |f(q,p)|,
\end{equation}
for instance.
Putting $c$ as
\begin{equation}
	c = \max\{c_1, c_2/\delta^2\}, 
\end{equation}
we obtain the inequality \eqref{eq:assumption-2}. $\blacksquare$

\section{Proofs of  Eqs.~\eqref{eq:int-avecos-zero}
  and \eqref{eq:int-p2cos-avecos2}}
\label{sec:appendix-proofs}

We set the considering integral as
\begin{equation}
    C = \iint \ave{\cos q}_{J} dqdp,
\end{equation}
and we will prove $C=0$.
We divide the $\mu$-space into two parts as
\begin{equation}
    X_{1} = \{(q,p) ~|~ k\leq k_{\ast} \},\quad
    X_{2} = \{(q,p) ~|~ k> k_{\ast} \},
\end{equation}
where $k$ is defined by \eqref{eq:k-p} and
\begin{equation}
    k_{\ast} = \sqrt{\dfrac{p_{\ast}^{2}}{4\Mb}+1}.
\end{equation}
Corresponding to this division of $\mu$-space, we divide $C$ as
\begin{equation}
    C = C_{1} + C_{2}
\end{equation}
where
\begin{equation}
    C_{1} = \iint_{X_{1}} \cos q~dqdp, 
    \quad
    C_{2} = \iint_{X_{2}} \ave{\cos q}_{J}dqdp.
\end{equation}
For the region $X_{1}$ we used Lemma \ref{lem:j-move} for $A=1$ and $B=\cos q$.
In the region $X_{2}$, $\ave{\cos q}_{J}$ is integrable
but $\cos q$ is not, and hence we cannot remove the average.
We separately compute $C_{1}$ and $C_{2}$.

For computing $C_{1}$, the recurrence relation
\begin{equation}
	\begin{split}
    	(2n+3)k^{-2} N[2n+4] - (2n+2)(1+k^{-2}) N[2n+2] \\+ (2n+1) N[2n] = 0
	\end{split}
\end{equation}
for the integrals
\begin{equation}
    N[2n] = \int_{0}^{1} \dfrac{u^{2n}}{\sqrt{(1-u^{2})(1-k^{-2}u^{2})}} du
\end{equation}
is useful, where the first two integrals are
\begin{equation}
    N[0] = K(1/k), \quad
    N[2] = k^{2}[ K(1/k) - E(1/k) ].
\end{equation}
The upper boundary of $X_{1}$ is expressed by
\begin{equation}
    p = b(q), \quad
    b(q) = 2\sqrt{\Mb} k_{\ast} \sqrt{1-k_{\ast}^{-2}\sin^{2}(q/2)},
\end{equation}
and hence the term $C_{1}$ is
\begin{equation}
    \begin{split}
        C_{1}
        & = \int_{-\pi}^{\pi} dq~\cos q \int_{-b(q)}^{b(q)} dp \\
        & = 8\sqrt{\Mb} k_{\ast} \int_{0}^{\pi} (1-2\sin^{2}(q/2))
        \sqrt{1-k_{\ast}^{-2}\sin^{2}(q/2)}dq \\
        & = 8\sqrt{\Mb} \left[
          -\dfrac{2k_{\ast}^{3}-k_{\ast}}{3} K(1/k_{\ast})
          + \dfrac{2k_{\ast}^{3}-k_{\ast}}{3} E(1/k_{\ast})
        \right].
    \end{split}
\end{equation}
In the way of computations, we used the change of variable as $u=\sin(q/2)$.

From the concrete form of $\ave{\cos q}_{J}$ for $k>1$, \eqref{eq:cosqJ-of-k},
the second term $C_{2}$ is directly computed as
\begin{equation}
    \begin{split}
        C_{2}
        = &4 \int_{k_{\ast}}^{\infty} dk \left(
          \dfrac{2k^{2}E(1/k)}{K(1/k)}-2k^{2}+1 \right)\\
        &\qquad\times
        \int_{0}^{\pi} dq \dfrac{2\sqrt{\Mb}}{\sqrt{1-k^{-2}\sin^{2}(q/2)}} \\
         = &8\sqrt{\Mb} \int_{k_{\ast}}^{\infty}
        \left[ (1-2k^{2})K(1/k) + 2k^{2}E(1/k) \right] dk.
    \end{split}
\end{equation}

Using the derivatives of $K(1/k)$ and $E(1/k)$,
\begin{equation}
    \begin{split}
        & \dfrac{\rm d}{{\rm d}k} K(1/k)
        = \dfrac{(1-k^{2})K(1/k)+k^{2}E(1/k)}{k(1-k^{2})}, \\
        & \dfrac{\rm d}{{\rm d}k} E(1/k)
        = \dfrac{K(1/k)-E(1/k)}{k},
    \end{split}
\end{equation}
we can show the equality
\begin{equation}
    \begin{split}
        & \dfrac{\rm d}{{\rm d}k} \left[ 
          - \dfrac{2(k^{3}-k)}{3} K(1/k)
          + \dfrac{2k^{3}-k}{3} E(1/k) \right] \\
        & = (1-2k^{2}) K(1/k) + 2k^{2}E(1/k).
    \end{split}
\end{equation}
This equality implies $C_{2}=-C_{1}$, and $C=0$ accordingly.
$\blacksquare$.

For the proof of \eqref{eq:int-p2cos-avecos2},
we set the integral of
\begin{equation}
    F = \iint_{\mu} \left( \dfrac{p^{2}}{2}\ave{\cos q}_{J} + \dfrac{\Mb}{4}
      - \Mb\ave{\cos q}_{J}^{2} \right) dqdp.
\end{equation}
We will prove $F=0$.
Using the relation \eqref{eq:k-p}, 
the equality \eqref{eq:int-avecos-zero} and Lemma \ref{lem:j-move},
we can modify $F$ as
\begin{equation}
    F = \Mb \iint_{\mu} \left( 2k^{2}\ave{\cos q}_{J} + \dfrac{1}{4} \right) dqdp.
\end{equation}
We divide the $\mu$-space into $X_{1}$ and $X_{2}$ again,
and $F$ into $F_{1}$ and $F_{2}$ accordingly, which are
\begin{equation}
    \begin{split}
        & F_{1} = \Mb \iint_{X_{1}} \left( 2k^{2}\cos q + \dfrac{1}{4} \right) dqdp\\
        & F_{2} = \Mb \iint_{X_{2}} \left( 2k^{2}\ave{\cos q}_{J} + \dfrac{1}{4} \right) dqdp.\\
    \end{split}
\end{equation}
We used the relation $\ave{k^{2}}_{J}=k^{2}$,
since $k$ depends on $(q,p)$ solely through the Hamiltonian $\HA$,
and the integral $\HA$ depends on $J$ only.
The remaining part of the proof can be performed
by a similar strategy with the proof of \eqref{eq:int-avecos-zero}
and we skip details.
$\blacksquare$


\end{document}